\documentclass[reqno,11pt,a4paper]{amsart}

\usepackage{amssymb}
\usepackage{url}
\usepackage{hyperref}
\usepackage{pdfpages}
\usepackage{graphicx}
\allowdisplaybreaks[4]

\usepackage{mathrsfs}
\let\mathcal\mathscr
\usepackage[mathscr]{eucal}
\usepackage[all]{xy}
\SelectTips{cm}{}

\usepackage{color}

\let\phi=\varphi
\let\kappa=\varkappa

\newcommand*{\ABC}{\textsc{ABC}}
\newcommand*{\Rom}[1]{\romannumeral #1}

\DeclareMathOperator{\const}{const}

\theoremstyle{theorem}

\theoremstyle{definition}

\theoremstyle{remark}
\newtheorem{remark}{Remark}

\usepackage[all]{xy}
\SelectTips{cm}{}

\usepackage{mathrsfs}
\let\mathcal\mathscr
\usepackage[mathscr]{eucal}

\makeatletter
\newcommand*{\rom}[1]{\expandafter\@slowromancap\romannumeral #1@}
\makeatother
\newcommand{\cprime}{\/{\mathsurround=0pt$'$}}

\begin{document}

\title[The \textsc{ABC}-equation revisited, {\rom{2}}]{The
  \textsc{ABC}-equation revisited: missed (quasi) integrable cases. Part
  \rom{2}: extensions and actions}
\author{I.S.~Krasil'shchik}
\address{Trapeznikov Institute of Control Sciences, 65 Profsoyuznaya street,
  Moscow 117997, Russia}
\email{josephkra@gmail.com}
\thanks{Partially supported by the RSF Grant 25-71-20008}
\begin{abstract}
  We continue here the study of the \textsc{ABC}-equation
  $Au_xu_{yt} + Bu_yu_{xt} + Cu_tu_{xy} = 0$, $A$, $B$, $C\in\mathbb{R}$ and
  using the results obtained in the first part construct new quasi-integrable
  structures and compute their actions on symmetries and cosymmetries.
\end{abstract} 
\date{\today}
\maketitle
\setcounter{tocdepth}{3}
\tableofcontents

\section{Intermezzo}
\label{sec:introduction}

In~\cite{ABC-1}, we studied the \ABC-equation
\begin{equation}
  \label{eq:1}
  Au_xu_{yt} + Bu_yu_{xt} + Cu_tu_{xy},\qquad A,B,C\in\mathbb{R},\quad
  A,B,C\neq 0,
\end{equation}
and found out that besides the well-known case (\Rom{1}) $A + B + C = 0$ there
also exist new ten ones:
(\Rom{2}) $A =  C$, $B =  C$;
(\Rom{3}) $A =  C$, $B = C$;
(\Rom{4}) $A =  C$, $2B =  3C$;
(\Rom{5}) $A =  C$, $B =  3C$;
(\Rom{6}) $A =  C$, $2B = 3C$;
(\Rom{7}) $A =  C$, $B = 3C$;
(\Rom{8}) $3A = 2C$, $3B =  2C$;
(\Rom{9}) $3A = C$, $3B =  C$;
(\Rom{10}) $A = C$, $2B =  C$;
(\Rom{11}) $A =  2C$, $B =  2C$. For these cases, nonlocal (quasi) integrable
structures were constructed. These structures are B\"{a}cklund transformations
that relate tangent and cotangent equations (see~\cite[Section~2.4]{ABC-1}).

We continue this study below and using the results obtained in~\cite{ABC-1}
and the extension procedure (Section~\ref{sec:comps-extens}) find new families
of nonlocal variables and new quasi-integrable structures. All them are
nonlocal. These results are presented in
Sections~\ref{sec:a-+-b}--\ref{sec:a-=-2c}.The above mentioned operators allow
one to claim the existence of infinite hierarchies of integrable structures;
this assertion is based on the composition of B\"{acklund} transformations. In
Section~\ref{sec:actions}, we present some results related to the actions of
integrable structures on nonlocal symmetries and cosymmetries. Finally, the
Appendices contain formulas that are too voluminous for the main text.

All the necessary definitions and constructions, as well as the relevant
references, can be found in~\cite[Section~2]{ABC-1}.

To enable computations, Equation~\eqref{eq:1} will be presented in the form
\begin{equation*}
  u_{xy} = \frac{au_xu_{yt} + bu_yu_{xt}}{u_t},
\end{equation*}
where $a = A/C$, $b = B/C$.

\section{Extensions}
\label{sec:comps-extens}

Recall that in all the B\"{a}cklund transformations constructed in~\cite{ABC-1} one of the
coverings was associated with a system of conservation laws
in~$\mathcal{T}\mathcal{E}$ or~$\mathcal{T}^*\mathcal{E}$; this covering will
be temporally denoted by~$\tau_1$. The secont covering was constructed by
means of a solution to equations~$\tilde{\ell}_{\mathcal{E}}(\mathcal{I}) = 0$
or~$\tilde{\ell}_{\mathcal{E}}^*(\mathcal{I}) = 0$ and will be denoted
by~$\tau_2$.

Let us describe the extension procedure now. Assume for example that
$\tau_1\colon W\to \mathcal{T}\mathcal{E}$ is a covering associated with a
conservation law~$\omega$ of~$\mathcal{T}\mathcal{E}$. Let~$\mathcal{I}$ be a
solution of the equation $\tilde{\ell}_{\mathcal{E}}(\mathcal{I}) = 0$ on~$W$
and $\tau_2\colon W\to \mathcal{T}\mathcal{E}$ be the corresponding
covering. If now $\tau_1'\colon W'\to \mathcal{T}\mathcal{E}$ is the covering
associated with another conservation law~$\omega'$, we can construct the
diagram
\begin{equation*}
  \xymatrix{
    W\times_{\mathcal{T}\mathcal{E}}
    W'\ar[r]\ar[d]_-{\tau_2^*(\tau_1')}&W'\ar[d]^-{\tau_1'}\\ 
    W\ar[r]^-{\tau_2}&\mathcal{T}\mathcal{E}
    \rlap{,}
  }
\end{equation*}
and $\tau_2'' = \tau_2\circ\tau_2^*(\tau_1')$ will be called the extension
of~$\tau_1'$ by meats of~$\tau_2$. In a similar way, if~$\mathcal{I}$ solves
the equation $\tilde{\ell}_{\mathcal{E}}^*(\mathcal{I}) = 0$ and~$\omega'$ is
a conservation law of~$\mathcal{T}^*\mathcal{E}$, we have the extension
\begin{equation*}
  \xymatrix{
    W\times_{\mathcal{T}^*\mathcal{E}}W'\ar[r]\ar[d]_-{\tau_2^*(\tau_1')}&W'\ar[d]^-{\tau_1'}\\
    W\ar[r]^-{\tau_2}&\mathcal{T}^*\mathcal{E}
  \rlap{.}}
\end{equation*}
These extensions are used in the constructions below.

\section{Cases}
\label{sec:cases}

\subsection{$A + B + C = 0$}
\label{sec:a-+-b}

There exist three recursion operators for symmetries
of~$\mathcal{E}$,~\cite[Section 3.1.1]{ABC-1}. Nine integrable structures
arise in the corresponding extension that relate symmetries and
cosymmetries. The simplest one are:
\begin{align*}
  \mathcal{I}_{11}
  &= \frac{u_t}{u_x^3}q + \frac{u_t^2}{f_3u_x^3}w_3 +
    \frac{2u_t}{af_1u_x^2}w_{13},\\ 
  \mathcal{I}_{21}
  &= -\frac{a(a - 1)f_2u_t}{2u_y^3}q - \frac{(a - 1)f_2u_xu_t}{f_1u_y^3}w_1 +
    \frac{u_t}{u_y^2}w_{21},\\
  \mathcal{I}_{31}
  &= -\frac{(a - 1)f_3}{u_t^2}q + \frac{f_3u_y}{f_2u_t^2}w_2 +
    \frac{1}{u_t}w_{32}. 
\end{align*}
The other six ones are given in Appendix~\ref{sec:a-+-b-1}. Here~$w_i$ are
nonlocal variables in~$\tau_2$, while~$w_{ij}$ are the variables in the
extension; they are defined in the same Appendix.

\subsection{$A =  -C$, $B =  -C$}
\label{sec:a-=-c}

In this case, the extended covering with the nonlocal variables~$w_{ij}$
(described in Appendix~\ref{sec:a-=-c-7}) admits three recursion operators for
symmetries. The simplest one is
\begin{multline*}
  \mathcal{R}_1 = 2f_3u_t^3q
  - \frac{f_3u_t^3(f_1u_x - f_{1, x}u)}{f_1^2u_x}w_{11}\\
  - \frac{f_3uu_t^3}{f_1u_x}w_{11,x}
  - \frac{2f_3u_t^3}{f_1u_x}w_{12} - \frac{f_3u_t^3}{f_2}w_{21} + u_tw_{31}.
\end{multline*}
The other two ones are also presented in Appendix~\ref{sec:a-=-c-7}.

\subsection{$A =  -C$, $B = C$}
\label{sec:a-=-c-1}

A sole recursion operator that arises in the extended covering is
\begin{multline*}
  \mathcal{R} = -\frac{f_1(f_{2, y}u + 4f_2u_y)u_x^3}{f_2u_y}q
  - \frac{f_1uu_x^3}{u_y}q_y + u_xw_{11}\\
  + \frac{f_1(f_2u_y - f_{2, y}u)u_x^3}{f_2^2u_y}w_{21}
  + \frac{f_1uu_x^3}{f_2u_y}w_{21,y}
  + \frac{f_1u_x^3}{f_2u_y}w_{22} + \frac{f_1u_x^3}{f_3}w_{31}.
\end{multline*}
The defining equations for~$w_{ij}$ are presented in
Appendix~\ref{sec:a-=-c-8}. 

\subsection{$A =  -C$, $2B =  -3C$}
\label{sec:a-=-c-2}

The equation admits two recursion operators:
\begin{align*}
  \mathcal{R}_1
  &= -u_t^3 q + \frac{u_t^3}{f_1\sqrt{u_x}}w_1
    + \frac{2u_t^3}{3f_2}w_{22} + \frac{u_t}{f_3}w_{31}
    \intertext{and}
    \mathcal{R}_2
  &= u_t^2 \left(\frac{9u_tu_{xt}}{8u_x} + \frac{u_tu_{yt}}{2u_y}
    - \frac{3u_{tt}}{2}\right)q
    - \frac{3u_t^4}{4u_x}q_x - \frac{u_t^4}{2u_y}q_y\\
  &- \frac{3u_t^2(u_tu_{xt} - 4u_xu_{tt})}{8f_1u_x^{3/2}}w_1
    + \frac{u_t^2u_{tt}}{f_2}w_{22}
    - \frac{3u_tf_{3, t}}{4f_3^2}w_{31} + \frac{3u_t}{4f_3}w_{31,t}.
\end{align*}
The defining equations for the nonlocal variables are given in
Appendix~\ref{sec:a-=-c-9}.

\subsection{$A =  -C$, $B =  -3C$}
\label{sec:a-=-c-3}

The equation possesses five recursion operators for symmetries:
\begin{align*}
  \mathcal{R}_0
  &= \frac{9f_1u_x^3}{2}q - \frac{9u_x}{2}w1 + u_xw_{11}
    + \frac{3 (f_3u_t^3 - f_1u_x^3)}{2f_2}w_{22} + u_tw_{31},\\
  \mathcal{R}_{11}
  &= \frac{1}{2f_1}\left(-\frac{u_x^3u_{xt}}{2u_t} + u_x^2u_{xx}
    - \frac{u_x^4u_{yt}}{6u_yu_t} + u_x^3f_{1, x}\right)q + u_x^3 q_x
    + \frac{u_x^4}{6u_y}q_y,\\
  &- \frac{u_x}{2f_1}w_{1,x} + \frac{u_x}{9f_1}w_{11,x}
    - \frac{u_x^2(f_{1, x}u_x + 2f_1u_{xx})}{6f_1f_2}w_{22},\\
  \mathcal{R}_{31}
  &= \frac{u_t^3u_{yt}}{u_y}q - \frac{u_t^4}{2u_y}q_y
    + \frac{u_t^2(f_{3, t}u_t + 2f_3u_{tt})}{2f_3f_2}w_{22}
    + \frac{u_t}{3f_3}w_{31,t}.
\end{align*}
The other two ones together with the description of the nonlocal variables are
presented in Appendix~\ref{sec:a-=-c-10}.

\subsection{$A =  -C$, $2B = 3C$}
\label{sec:a-=-c-4}

There exist three recursion operators for symmetries of this equation. One
of them is
\begin{equation*}
  \mathcal{R}_0 = -2u_x^3q + \frac{8u_x}{9f_1}w_{11}
  + \frac{2u_x^3}{3f_2}w_{22} + \frac{u_x^3}{f_3\sqrt{u_t}}w_3.
\end{equation*}
The operators~$\mathcal{R}_1$ and~~$\mathcal{R}_2$ see in
Appendix~\ref{sec:a-=-c-11}. Defining equations for $w$'s are in the same
Appendix.

\subsection{$A =  -C$, $B = 3C$}
\label{sec:a-=-c-5}

The equation admits six recursion operator including
\begin{align*}
  \mathcal{R}_{11}
  &= -\frac{9f_1u_x^3}{2}q + u_xw_{11} + \frac{3f_1u_x^3}{2f_2}w_{22},\\
  \mathcal{R}_{31}
  &= -2f_3u_t^3q + \frac{f_3u_t^3}{3f_2}w_{22} + u_tw_3
    - \frac{2u_t}{9}w_{31}
\end{align*}
and other four presented, together with the defining equations for the
nonlocal variables, in Appendix~\ref{sec:a-=-c-12}. It also possesses the
integrable structure
\begin{equation*}
  \mathcal{I} =  -\frac{9u_t q}{2} + \frac{u_tw_{21}}{f_2u_y}
  - \frac{(f_{2, y} u - 2f_2u_y) u_t w_{22}}{2f_2^2u_y}
  + \frac{uu_tw_{22,y}}{2f_2u_y},
\end{equation*}
where
\begin{align*}
  w_{21,x}
  &= f_2 \Big(-\frac{9uu_yu_{xtt}}{2u_t^2} -\frac{uu_xu_{yyt}}{2u_tu_y}
    - \frac{3(2uu_tu_{yt} - 3uu_yu_{tt} + u_t^2u_y) u_{xt}}{2u_t^3}\\
  &+ \frac{u_x(uu_t^2u_{yy} - 3uu_y^2u_{tt}  + u_t^2u_y^2)u_{yt}}{2 u_t^3
    u_y^2}
    + \frac{3uu_xu_{ytt}}{2u_t^2}\Big) q
    + \frac{3f_2(u_tu_y-uu_{yt}) q_x}{2u_t}\\
  &- \frac{f_2(uu_tu_xu_{yy} - 9uu_y^2u_{xt}  + u_tu_xu_y^2)q_y}{2u_tu_y^2}
    + \frac{3f_2 u (u_x u_{yt} - 3u_yu_{xt}) q_t}{2u_t^2}\\
  &+ \frac{f_2uu_xq_{yy}}{2u_y} + \frac{9f_2uu_yq_{xt}}{2u_t}
    - \frac{3f_2 u u_x q_{yt}}{2 u_t},\\
  w_{21,t}
  &= \frac{f_2(u u_{yt} u_{yy} + 4 u_y^2 u_{yt}-uu_yu_{yyt}) q}{2u_y^2}\\
  &- \frac{f_2u_t(u u_{yy} + 4 u_y^2) q_y}{2 u_y^2}
    + \frac{f_2 u u_t q_{yy}}{2 u_y};\\
    w_{22,x}
  &= \frac{f_2(u_xu_{yt} - 3u_yu_{xt}) q}{u_yu_t}
    + \frac{3 f_2(q_x - u_x q_y)}{u_y},\\
  w_{22,t}
  &= \frac{f_2 u_{yt} q}{u_y} - \frac{f_2 u_t q_y}{u_y}.
\end{align*}

\subsection{$3A = 2C$, $3B =  -2C$}
\label{sec:3a-=-2c}

There are three recursion operators for symmetries of this equation. The first
one is
\begin{equation*}
  \mathcal{R}_1 = -\frac{3f_2u_y^3}{2f_1\sqrt{u_x}}w_1 + u_yw_{21}
  + \frac{f_2u_y^3}{f_3}w_{31}.
\end{equation*}
The other two and the definitions of the nonlocal variables see in Appendix~\ref{sec:3a-=-2c-1}.

\subsection{$3A = - C$, $3B =  C$}
\label{sec:3a-=-c}

The equation admits six recursion operators for symmetries in this case. The
simplest two ones are
\begin{align*}
  \mathcal{R}_{11}
  &= u_xw_{11}
    - \frac{f_1(f_3u_t + f_{3, t}u)u_x^3}{3f_3^2u_t}w_{31} +
    \frac{f_1uu_x^3}{3f_3u_t}w_{31,t}
    + \frac{2f_1u_x^3}{3f_3u_t}w_{32},\\
  \mathcal{R}_{21}
  &= 2f_2u_y^3q - 3u_yw_2 + u_yw_{21} - \frac{f_2u_y^3}{f_3}w_{31}.
\end{align*}
The other four and the description of nonlocal variables are given in
Appendix~\ref{sec:3a-=-c-1}.

\subsection{$A = C$, $2B =  -C$}
\label{sec:a-=-c-6}

We found the following recursion operators for symmetries in this case:
\begin{align*}
  \mathcal{R}_1
  &= \frac{u_y^3}{f_1\sqrt{u_x}}w_1 + \frac{2 u_y}{f_2}w_{21},\\
  \mathcal{R}_2
  &= -\frac{u_y^3(2u_xu_{yt} - u_yu_{xt})}{2u_tu_x}q
    - \frac{u_y^4}{u_x}q_x
    \\
  &+ u_y^2 \left(\frac{4u_{yy}}{f_1\sqrt{u_x}}
    - \frac{u_y^2u_{xt}}{2u_x^{3/2}u_t}
    + \frac{u_yu_{yt}}{f_1u_t\sqrt{u_x}}\right)w_1- \frac{4f_{2, y}u_y}{f_2^2}w_{21} + \frac{4u_y}{f_2}w_{21,y},
\end{align*}
and the third one which is presented, together with the description of the
nonlocal variables, in Appendix~\ref{sec:a-=-c-13}.

We also found an integrable structure of the form
\begin{equation*}
  \mathcal{I} = \frac{1}{f_1\sqrt{u_x}}w_{22} + \frac{1}{f_3\sqrt{u_t}}w_{23}
  + \frac{1}{u_t}p
\end{equation*}
whose action is inverse to the one presented in~\cite[Section~3.11]{ABC-1}. The
variables~$w_{22}$ and~$w_{23}$ are also described in
Appendix~\ref{sec:a-=-c-13}.

\subsection{$A =  -2C$, $B =  -2C$}
\label{sec:a-=-2c}

The equation admits three recursion operators for symmetries. The first one is
\begin{equation*}
  \mathcal{R}_1 = f_3u_t^3 q - \frac{f_3u_t^3}{2f_1\sqrt{u_x}}w_1 + u_tw_{31}.
\end{equation*}
The other two one can find in Appendix~\ref{sec:a-=-2c-1}.

There also exists an integrable structure
\begin{equation*}
  \mathcal{I} = \frac{1}{f_1\sqrt{u_x}}w_{33} + \frac{2}{u_t}p
  + \frac{1}{f_2\sqrt{u_y}}w_{32}
\end{equation*}
that relates cosymmetries and symmetries.

All the nonlocal variables are described in Appendix~\ref{sec:a-=-2c-1}.

\section{Actions}
\label{sec:actions}

We describe here the actions of the above described structures on symmetries
and cosymmetries of the \ABC-equation. The case~$A = B = C$ is considered in
detail (cf.~\cite{Veronese-web}); in all the other cases we mainly pay
attention to the nonlocal objects arising under the actions. The results of
actions are presented modulo the image of zero. The nonlocal
variables~$z_\sigma$ associated with the two-component conservation laws of
the \ABC-equation are described in Appendix~\ref{sec:nonl-vari-abc}.

\subsection{$A + B + C = 0$}
\label{sec:=-b-=}

Recall that the equation admits the following recurion operators in this
case,~\cite[Sec.~3.1.1]{ABC-1}: 
\begin{equation*}
  \mathcal{R}_1 = \frac{u_xw_1}{f_1},\quad
  \mathcal{R}_2 = \frac{u_yw_2}{f_2},\quad
  \mathcal{R}_3 = \frac{u_tw_3}{f_3},
\end{equation*}
where $w_i$ are nonlocal variables in~$\mathcal{T}(\mathcal{E})$ also
described in,~\cite{ABC-1}. The action of~$\mathcal{R}_1$ reads
\begin{gather*}
  \mathcal{R}_1(0) = \phi_1,\quad
  \mathcal{R}_1(\phi_0) = -\frac{(a - 1) \phi_0}{2},\\
  \mathcal{R}_1(\phi_1)
  = \frac{u_x h_1 z_{1,x}}{2f_1}
  - \frac{u_x (h_1 f_1' + f_1 h_1') z_1}{2 f_1^2},\\
  \mathcal{R}_1(\phi_2) = 0,\quad \mathcal{R}_1(\phi_3) = -\frac{a\phi_3}{2},
\end{gather*}
while the inverse one is
\begin{gather*}
  \mathcal{R}_1(0) = \phi_2,\quad
  \mathcal{R}_1(\phi_0) = -\frac{2 \phi_0}{a - 1},\quad
  \mathcal{R}_1(\phi_1) = 0,\\
  \mathcal{R}_1(\phi_2) = \frac{2h_2 u_y z_{2,y}}{a(a - 1)f_2}
  -\frac{2(h_2' f_2 + f_2' h_2) u_y z_2}{a(a - 1)f_2^2}
  ,\quad
  \mathcal{R}_1(\phi_3) = -\frac{2\phi_3}{a}.
\end{gather*}
Thus, we have the following diagram
\begin{equation*}
  \xymatrix{
    \dots\ar@<.5ex>[r]^-{\mathcal{R}_1}
    &\dfrac{\Phi_2^1}{a(a-1)}\ar@<1ex>[l]^-{\mathcal{R}_1}\ar@<.5ex>[r]^-{\mathcal{R}_1}
    &\phi_2\ar@<.5ex>[r]^-{\mathcal{R}_1}\ar@<1ex>[l]^-{\mathcal{R}_1}
    &0\ar@<.5ex>[r]^-{\mathcal{R}_1}\ar@<1ex>[l]^-{\mathcal{R}_1}
    &\phi_1\ar@<.5ex>[r]^-{\mathcal{R}_1}\ar@<1ex>[l]^-{\mathcal{R}_1}
    &\Phi_1^1\ar@<.5ex>[r]^-{\mathcal{R}_1}\ar@<1ex>[l]^-{\mathcal{R}_1}
    &\ar@<1ex>[l]^-{\mathcal{R}_1}\dots\rlap{,}
    \\
    &\phi_0\ar@<.5ex>[r]^-{\mathcal{R}_1}
    &\ar@<1ex>[l]^-{\mathcal{R}_1}\phi_0\rlap{,}&
    &\phi_3\ar@<.5ex>[r]^-{\mathcal{R}_1}
    &\phi_3\ar@<1ex>[l]^-{\mathcal{R}_1}\rlap{.}&
  }
\end{equation*}
In a similar way,
\begin{equation*}
  \xymatrix{
    \dots\ar@<.5ex>[r]^-{\mathcal{R}_2}
    &\dfrac{\Phi_3^1}{a-1}\ar@<1ex>[l]^-{\mathcal{R}_2}\ar@<.5ex>[r]^-{\mathcal{R}_2}
    &\phi_3\ar@<.5ex>[r]^-{\mathcal{R}_2}\ar@<1ex>[l]^-{\mathcal{R}_2}
    &0\ar@<.5ex>[r]^-{\mathcal{R}_2}\ar@<1ex>[l]^-{\mathcal{R}_2}
    &\phi_2\ar@<.5ex>[r]^-{\mathcal{R}_2}\ar@<1ex>[l]^-{\mathcal{R}_2}
    &\Phi_2^1\ar@<.5ex>[r]^-{\mathcal{R}_2}\ar@<1ex>[l]^-{\mathcal{R}_2}
    &\ar@<1ex>[l]^-{\mathcal{R}_2}\dots\rlap{,}
    \\
    &\phi_0\ar@<.5ex>[r]^-{\mathcal{R}_2}
    &\ar@<1ex>[l]^-{\mathcal{R}_2}\phi_0\rlap{,}&
    &\phi_1\ar@<.5ex>[r]^-{\mathcal{R}_2}
    &\phi_1\ar@<1ex>[l]^-{\mathcal{R}_2}\rlap{,}&
  }
\end{equation*}
and
\begin{equation*}
  \xymatrix{
    \dots\ar@<.5ex>[r]^-{\mathcal{R}_3}
    &\dfrac{\Phi_1^1}{a}\ar@<1ex>[l]^-{\mathcal{R}_3}\ar@<.5ex>[r]^-{\mathcal{R}_3}
    &\phi_1\ar@<.5ex>[r]^-{\mathcal{R}_3}\ar@<1ex>[l]^-{\mathcal{R}_3}
    &0\ar@<.5ex>[r]^-{\mathcal{R}_3}\ar@<1ex>[l]^-{\mathcal{R}_3}
    &\phi_3\ar@<.5ex>[r]^-{\mathcal{R}_3}\ar@<1ex>[l]^-{\mathcal{R}_3}
    &\Phi_3^1\ar@<.5ex>[r]^-{\mathcal{R}_3}\ar@<1ex>[l]^-{\mathcal{R}_3}
    &\ar@<1ex>[l]^-{\mathcal{R}_3}\dots\rlap{,}
    \\
    &\phi_0\ar@<.5ex>[r]^-{\mathcal{R}_3}
    &\ar@<1ex>[l]^-{\mathcal{R}_3}\phi_0\rlap{,}&
    &\phi_2\ar@<.5ex>[r]^-{\mathcal{R}_3}
    &\phi_2\ar@<1ex>[l]^-{\mathcal{R}_3}\rlap{,}&
  }
\end{equation*}
where
\begin{gather*}
  \Phi_1^1 =  \frac{h_1u_xz_{1,x}}{f_1} - \frac{u_x(h_1f_1' +
    f_1h_1')z_1}{f_1^2},\\ 
  \Phi_2^1 = \frac{h_2u_yz_{2,y}}{f_2} - \frac{u_y(f_2h_2' +
    h_2f_2')z_2}{f_2^2},\\ 
  \Phi_3^1= \frac{h_3u_tz_{3,t}}{f_3} - \frac{u_t(h_3f_3' +
    f_3h_3')z_3}{f_3^2}. 
\end{gather*}
are nonlocal symmetries.

The action of recursion operators
\begin{equation*}
  \mathcal{R}_1 = \frac{u_t w_1}{f_1u_x^2},\quad
  \mathcal{R}_2 = \frac{u_t w_2}{f_2u_y^2},\quad
  \mathcal{R}_3 = \frac{w_3}{f_3 u_t}
\end{equation*}
on the first-order cosymmetries has a similar form (actually, it suffices to
change all the~$\phi$'s and~$\Phi$'s to~$\psi$'s and~$\Psi$'s in the formulas
above), but the action on the third-order
elements~$\psi_{41},\dots,\psi_{47}$ is a bit more complicated. For example,
one has for~$\mathcal{R}_1$
\begin{gather*}
  \mathcal{R}_1(\psi_{41}) = -(a - 1) \psi_{44},\quad
  \mathcal{R}_1(\psi_{42}) = (a - 1) ((a - 1) \psi_{44} - a \psi_{41}),\\
  \mathcal{R}_1(\psi_{43}) = \psi_{46} - a \psi_{43},\quad
  \mathcal{R}_1(\psi_{44}) = \psi_{43} - a \psi_{44},\\
  \mathcal{R}_1(\psi_{45}) = (a - 1) (a (a - 1) \psi_{41} - (a - 1)^2
  \psi_{44} - a \psi_{42}),\\ 
  \mathcal{R}_1(\psi_{46}) = -(a - 1) \psi_{46} + \Psi_{13},\quad
  \mathcal{R}_1(\psi_{47}) = \psi_{44} - a \psi_{47},
\end{gather*}
while the action of the inverse operator is given by
\begin{gather*}
  \mathcal{R}_1(\psi_{41}) = -\frac{\psi_{42}}{a(a - 1)} -
  \frac{\psi_{41}}{a},\quad 
  \mathcal{R}_1(\psi_{42}) = -\frac{\psi_{42}}{a} - \frac{\psi_{45}}{a (a - 1)},\\
  \mathcal{R}_1(\psi_{43}) = \psi_{44} - \frac{a\psi_{41}}{a - 1}, \quad
  \mathcal{R}_1(\psi_{44}) = -\frac{\psi_{41}}{a - 1},\quad
  \mathcal{R}_1(\psi_{45}) = -\frac{\psi_{45}}{a - 1} + \Psi_{32},\\
  \mathcal{R}_1(\psi_{46}) = \psi_{43} + a\psi_{44} - \frac{a^2\psi_{41}}{a -
    1},\quad 
  \mathcal{R}_1(\psi_{47}) = -\frac{\psi_{47}}{a} - \frac{\psi_{41}}{a (a - 1)}.
\end{gather*}
We have the
\begin{equation*}\xymatrixcolsep{8mm}
  \xymatrix{
    \dots\ar@<.5ex>[r]
    &\Psi_{23}\ar@<.5ex>[r]\ar@<1ex>[l]
    &\psi_{45}\ar@<.5ex>[r]\ar@<1ex>[l]
    &\bar{\psi}_{44}\ar@<.5ex>[r]\ar@<1ex>[l]
    &\bar{\psi}_{43}\ar@<.5ex>[r]\ar@<1ex>[l]
    &\bar{\psi}_{46}\ar@<.5ex>[r]\ar@<1ex>[l]
    &\bar{\Psi}_{13}\ar@<.5ex>[r]\ar@<1ex>[l]
    &\dots\ar@<1ex>[l]\rlap{,}
  }
\end{equation*}
where
\begin{gather*}
  \bar{\psi}_{44} = (a - 1)\big(a(a - 1)\psi_{41} - (a - 1)^2\psi_{44} -
  a\psi_{42}\big),\\ 
\bar{\psi}_{43} = (a - 1)^2 \big(a^2 \psi_{41} + (-a + 1) \psi_{43} - a (a - 1)
\psi_{44}\big),\\ 
\bar{\psi}_{46} = -(a - 1)^3\psi_{46},\quad 
\bar{\Psi}_{13} = (a - 1)^3 \big((a - 1) \psi_{46} - \Psi_{13}\big).
\end{gather*}
The actions of~$\mathcal{R}_2$ and~$\mathcal{R}_3$ are similar.

Note that the recursion operators give rise to the following nonlocal
cosymmetries
\begin{align*}
  \Psi_{13}
  &= \frac{u_t (6f_1''f_1'f_1 - 6f_1'^3 - f_1'''f_1^2)z_1}{f_1^4u_x^2}
    + \frac{3u_t(2(f_1'^2 - f_1''f_1) z_{1,x}}{f_1^3u_x^2}\\
  &- \frac{3 u_tf_1'z_{1,xx}}{f_1^2u_x^2} + \frac{u_tz_{1,xxx}}{f_1u_x^2},\\
  \Psi_{23}
  &= -\frac{u_t(f_2'''f_2^2 + 6f_2'^3 - 6f_2'f_2''f_2)z_2}{f_2^4u_y^2}
    + \frac{3u_t(2f_2'^2 - f_2''f_2)z_{2,y}}{f_2^3u_y^2}\\
  &- \frac{3u_tf_2'z_{2,yy}}{f_2^2 u_y^2}
    + \frac{u_t z_{2,yyy}}{f_2 u_y^2},\\
  \Psi_{33}
  &= \frac{(6f_3'^3 - 6f_3''f_3'f_3 + f_3'''f_3^2)z_3}{f_3^4u_t}
    - \frac{3(2f_3'^2 - f_3''f_3)z_{3,t}}{f_3^3u_t}\\
  &+ \frac{3f_3'z_{3,tt}}{f_3^2u_t} - \frac{z_{3,ttt}}{f_3u_t}.
\end{align*}

\subsection{$A = -C$, $B = -C$}
\label{sec:a-=-c-14}

The recursion operator gives rise to the nonlocal symmetries
\begin{equation*}
  \Phi_3^0 = \frac{u_tz_3}{f_3},\qquad \Phi_3^1 = \frac{u_tz_{3, t}}{f_3}
\end{equation*}
and to the diagram
\begin{equation*}
  \xymatrix{
    \dots\ar@<.5ex>[r]^-{\mathcal{R}}
    &\Phi_3^0\ar@<.5ex>[r]^-{\mathcal{R}}\ar@<.5ex>[l]^-{\mathcal{R}}
    &0\ar@<.5ex>[r]^-{\mathcal{R}}\ar@<.5ex>[l]^-{\mathcal{R}}
    &\phi_3\ar@<.5ex>[r]^-{\mathcal{R}}\ar@<.5ex>[l]^-{\mathcal{R}}
    &\Phi_3^1\ar@<.5ex>[r]^-{\mathcal{R}}\ar@<.5ex>[l]^-{\mathcal{R}}
    &\dots\ar@<.5ex>[l]^-{\mathcal{R}}
  }
\end{equation*}
The integrable structure~$\mathcal{I}$ from~\cite[Sec.~3.2]{ABC-1} generates
the symmetry~$\Phi_3^1$, while the inverse action gives four nonlocal
cosymmetries. The simplest one is
\begin{equation*}
  \Psi_1 = -\frac{f_{1,x}u_tz_{12}}{u_xf_1^2}
  + \frac{u_tz_{12,x}}{f_1u_x}
  - \frac{f_{2,y}u_tz_{21}}{f_2^2u_y}
  + \frac{u_tz_{21,y}}{f_2u_y} - (\ln(u_x) + \ln(u_y) + 2) u_t;
\end{equation*}
the other three are presented in Appendix~\ref{sec:case-a-=}.

\subsection{$A = -C$, $B = C$}
\label{sec:a-=-c-15}

In this case, the recursion operator generates the nonlocal symmetries
\begin{gather*}
  \Phi_0^1 = h_1u_xz_1,\qquad
  \Phi_1^1 = \frac{u_xz_{1,x}}{f_1},\\
  \Phi_3^1 =
  \frac{h_3u_t(f_3^2u_t\ln(u_t) - f_3^2u_t - f_3'z_{31} + f_3z_{31,t})}{f_3^2}
\end{gather*}
and the sequence
\begin{equation*}
  \xymatrix{
    \dots\ar@<.5ex>[r]^-{\mathcal{R}}
    &\ar@<.5ex>[r]^-{\mathcal{R}}\ar@<.5ex>[l]^-{\mathcal{R}}\Phi
    &\ar@<.5ex>[r]^-{\mathcal{R}}\ar@<.5ex>[l]^-{\mathcal{R}}0
    &\ar[r]^-{\mathcal{R}}\ar@<.5ex>[l]^-{\mathcal{R}}\phi_1&0\rlap{,}
  }
\end{equation*}
where~$\Phi = \Phi_0^1 + \Phi_1^1 + \Phi_3^1$.

The inverse action of the integrable structure~$\mathcal{I}$ produces the
symmetry~$\Phi_0^1$, while the direct one gives rize to four nonlocal
cosymmetries presented in Appendix~\ref{sec:nonl-cosymm-case}.

\subsection{$A = -C$, $2B = -3C$}
\label{sec:a-=-c-16}

Both recursion operators give rise to the following sequence
\begin{equation*}\xymatrixcolsep{12mm}
  \xymatrix{
    \dots\ar@<.5ex>[r]^-{\mathcal{R}_1, \mathcal{R}_2}
    &\ar@<.5ex>[r]^-{\mathcal{R}_1, \mathcal{R}_2}\ar@<.5ex>[l]^-{\mathcal{R}_1, \mathcal{R}_2}
    \Phi_2^1 + \Phi_3^1
    &\ar@<.5ex>[r]^-{\mathcal{R}_1, \mathcal{R}_2}\ar@<.5ex>[l]^-{\mathcal{R}_1, \mathcal{R}_2}0
    &\ar[r]^-{\mathcal{R}_1, \mathcal{R}_2}\ar@<.5ex>[l]^-{\mathcal{R}_1, \mathcal{R}_2}\phi_3&0\rlap{,}
  }
\end{equation*}
where
\begin{gather*}
  \Phi_2^1 =
  \frac{h_2u_y(4f_2^2u_y\ln(u_y) + f_2^2u_y  - 4f_2'z_{22} + 4f_2z_{22,y})}{3f_2^2} - \frac{2h_2'u u_y}{3},\\
  \Phi_3^1 =
  \frac{h_3(4f_3u_t^{3/2}z_3 + f_3u_t^{3/2}z_{3,t} - 2f_3^2)}{f_3\sqrt{u_t}}
  + 5h_3'u_t z_3.
\end{gather*}
The integrable structure~$\mathcal{I}$ generates four nonlocal
cosymmetries. The simplest two are
\begin{align*}
  \Psi_1^1
  &= -\frac{(2g_1f_1' - f_1g_1')u_tz_1}{2f_1^2\sqrt{u_x}} +
    \frac{g_1u_tz_{1,x}}{f_1\sqrt{u_x}},\\
  \Psi_2^1
  &= \frac{4f_2'u_tz_{22}}{f_2^2u_y}
    - \frac{4u_tz_{22,y}}{f_2u_y} + 4 (\ln(u_y) + 5) u_t.
\end{align*}
The other two are presented in Appendix~\ref{sec:nonl-cosymm-a}.

\subsection{$A = -C$, $B = -3C$}
\label{sec:a-=-c-17}

Recall that the equation admits five recursion operators in this case. Their
action is as follows:
\begin{equation*}\xymatrixcolsep{12mm}
  \xymatrix{
    \dots\ar@<.5ex>[r]^-{\mathcal{R}_0}
    &\ar@<.5ex>[r]^-{\mathcal{R}_0}\ar@<.5ex>[l]^-{\mathcal{R}_0}
    \Phi_3
    &\ar@<.5ex>[r]^-{\mathcal{R}_0}\ar@<.5ex>[l]^-{\mathcal{R}_0}0
    &\ar[r]^-{\mathcal{R}_0}\ar@<.5ex>[l]^-{\mathcal{R}_0}\phi_1+\phi_3
    &0\rlap{,}\\
    \dots\ar@<.5ex>[r]^-{\mathcal{R}_{11}, \mathcal{R}_{12}}
    &\ar@<.5ex>[r]^-{\mathcal{R}_{11}, \mathcal{R}_{12}}\ar@<.5ex>[l]^-{\mathcal{R}_{11}, \mathcal{R}_{12}}
    \Phi_2 + \Phi_3
    &\ar@<.5ex>[r]^-{\mathcal{R}_{11}, \mathcal{R}_{12}}\ar@<.5ex>[l]^-{\mathcal{R}_1, \mathcal{R}_2}0
    &\ar[r]^-{\mathcal{R}_{11}, \mathcal{R}_{12}}\ar@<.5ex>[l]^-{\mathcal{R}_{11},
      \mathcal{R}_{12}}\phi_1
    &0\rlap{,}\\
    \dots\ar@<.5ex>[r]^-{\mathcal{R}_{31}, \mathcal{R}_{32}}
    &\ar@<.5ex>[r]^-{\mathcal{R}_{31}, \mathcal{R}_{32}}\ar@<.5ex>[l]^-{\mathcal{R}_{31}, \mathcal{R}_{32}}
    \Phi_1 + \Phi_2 + \Phi_3
    &\ar@<.5ex>[r]^-{\mathcal{R}_{31}, \mathcal{R}_{32}}\ar@<.5ex>[l]^-{\mathcal{R}_{31}, \mathcal{R}_{32}}0
    &\ar[r]^-{\mathcal{R}_{31}, \mathcal{R}_{32}}\ar@<.5ex>[l]^-{\mathcal{R}_{31}, \mathcal{R}_{32}}\phi_3&0\rlap{,}
  }
\end{equation*}
where
\begin{align*}
  \Phi_1
  &=\frac{u_x(2f_1^2u_x^3  + 2f_1z_{1,x} - f_{1, x}z_1)h_1}{f_1}
    + h_{1, x}u_xz_1,\\
  \Phi_2
  &= \frac{u_y(f_2^2u_y\ln(u_y) - 2f_2^2u_y + f_2z_{22,y} -
    f_{2,y}z_{22})h_2}{f_2^2} 
    - \frac{u u_y h_{2, y}}{2},\\
  \Phi_3
  &= \frac{(f_3u_t^3z_{3t} - f_{3,t}u_t^3z_3 - f_3^2) h_3}{2f_3^2u_t^2}
    + \frac{h_{3,t}u_tz_3}{f_3}.
\end{align*}
Actions of the integrable structure~$\mathcal{I}$ on local symmetries give
rise to the nonlocal cosymmetries
\begin{align*}
  \Psi_1
  &= \frac{f_{2, y}u_tz_{22}}{f_2^2u_y}
    - \frac{u_tz_{22,y}}{f_2u_y} + u_t(\ln(u_y) + 2),\\
  \Psi_2
  &= \frac{2u_tz_{21}}{f_2u_y}
    - \frac{u_t(f_{2, y}u - 2f_2u_y) z_{22}}{2f_2^2u_y}
    + \frac{uu_tz_{22,y}}{2f_2u_y} + \frac{u_t (\ln(u_y) u - 6 u)}{2},\\
  \Psi_3
  &= -\frac{u_t(f_{2,y}g_2 - g_{2,y})f_2) z_{21}}{f_2^2u_y}\\
  & + \frac{u_t(2f_{2, y}^2g_2u - g_{2, y}f_{2, y}f_2u
    - 2f_{2, y}g_2f_2u_y - f_{2, y, y}f_2 g_2 u) z_{22}}{2f_2^3u_y}\\
  &- \frac{u_t(2f_{2, y}g_2u - g_{2, y}f_2u - 2g_2f_2u_y)z_{22,y}}{2f_2^2u_y}
    + \frac{g_2uu_tz_{22,yy}}{2f_2u_y}\\
  &+ \frac{g_2u_tz_{21,y}}{f_2u_y} + \frac{g_2u_t(\ln(u_y - 4)u_y}{2}
    - g_{2, y}uu_t  - \frac{g_{2, y, y}u^2u_t}{4u_y}.
\end{align*}

\subsection{$A = -C$, $2B = 3C$}
\label{sec:a-=-c-18}

The action of the operator~$\mathcal{R}_0$ is
\begin{equation*}
  \xymatrix{\dots\ar@<.5ex>[r]^-{\mathcal{R}_0}
    &\ar@<.5ex>[r]^-{\mathcal{R}_0}\ar@<.5ex>[l]^-{\mathcal{R}_0}
    \Phi_{01} + \Phi_{02}
    &\ar@<.5ex>[r]^-{\mathcal{R}_0}\ar@<.5ex>[l]^-{\mathcal{R}_0}0
    &\ar[r]^-{\mathcal{R}_0}\ar@<.5ex>[l]^-{\mathcal{R}_0}\phi_1
    &0\rlap{,}
  }
\end{equation*}
where
\begin{align*}
  \Phi_{01}
  &= \frac{2(4f_{1, x}u_x^{3/2}z_1 + f_1u_x^{3/2}z_{1,x} -
            2 f_1^2) h_1}{3f_1\sqrt{u_x}} + \frac{10h_{1, x}u_xz_1}{3},\\
  \Phi_{02}
  &= \frac{u_y(4f_2^2u_y(\ln(u_y) + u_y) - 2f_{2,y}(f_2u + 2z_{22})
    + 4f_2z_{22,y}) h_2}{3f_2} - \frac{2h_{2, y}f_2uu_y}{3}.
\end{align*}

The inverse action of the integrable structure~$\mathcal{J}$ gives rise to the
nonlocal symmetry
\begin{equation*}
  \Phi = \frac{h_1(f_1 u_x^{3/2}z_{1,x} - f_{1, x}u_x^{3/2}z_1 - 2 f_1^2)}{5f_1^2\sqrt{u_x}} + \frac{h_{1, x}u_x z_1}{f_1}.
\end{equation*}

\subsection{$A = -C$, $B = 3C$ }
\label{sec:a-=-c-19}

Let us describe here the action of~$\mathcal{R}_{11}$
znd~$\mathcal{R}_{31}$. One has
\begin{gather*}
  \xymatrix{\dots\ar@<.5ex>[r]^-{\mathcal{R}_{11}}
    &\ar@<.5ex>[r]^-{\mathcal{R}_{11}}\ar@<.5ex>[l]^-{\mathcal{R}_{11}}
    \Phi_1 + \Phi_2
    &\ar@<.5ex>[r]^-{\mathcal{R}_{11}}\ar@<.5ex>[l]^-{\mathcal{R}_{11}}0
    &\ar[r]^-{\mathcal{R}_{11}}\ar@<.5ex>[l]^-{\mathcal{R}_{11}}\phi_1
    &0\rlap{,}}\\
  \xymatrix{\dots\ar@<.5ex>[r]^-{\mathcal{R}_{31}}
    &\ar@<.5ex>[r]^-{\mathcal{R}_{31}}\ar@<.5ex>[l]^-{\mathcal{R}_{31}}
    \Phi_1 + \Phi_2
    &\ar@<.5ex>[r]^-{\mathcal{R}_{31}}\ar@<.5ex>[l]^-{\mathcal{R}_{31}}0
    &\ar[r]^-{\mathcal{R}_{31}}\ar@<.5ex>[l]^-{\mathcal{R}_{31}}\phi_3
    &0\rlap{,}
  }
\end{gather*}
where
\begin{align*}
  \Phi_1
  &= \frac{h_1(f_{1, x}u_x^3 z_1 + f_1u_x^3 z_{1,x} - f_1^2)}{3f_1u_x^2} +
         \frac{2h_{1, x}u_xz_1}{3},\\
  \Phi_2
  &= \frac{h_2u_y(2f_2^2u_y(\ln u_y + 1) - f_{2, y} u f_2 + 2 z_{22,y} f_2 - 2f_{2, y}z_{22})}{3f_2} - \frac{f_2h_{2, y}uu_y}{3}.
\end{align*}

The direct action of the structure~$\mathcal{J}$ from
Section~\ref{sec:a-=-c-5} gives rise to the following three nonlocal
cosymmetries:
\begin{align*}
  \Psi_1
  &=\frac{f_{2, y} u_t z_{22}}{2f_2^2u_y}
    - \frac{u_tz_{22,y}}{2f_2u_y} + \frac{(\ln(u_y) - 7) u_t}{2},\\
  \Psi_2
  &= \frac{2 u_t z_{21}}{f_2 u_y}
    + \frac{u_t (-f_{2, y} u + 2 f_2 u_y) z_{22}}{2f_2^2u_y}
    + \frac{uu_tz_{22,y}}{2f_2u_y} + \frac{uu_t(\ln(u_y) - 3)}{2},\\
  \Psi_3
  &= -\frac{u_t(f_{2, y}h_2 - f_2h_{2, y}) z_{21}}{f_2^2u_y}
    + \frac{h_2u_tz_{21,y}}{f_2u_y}\\
  &- u_t \frac{(f_{2, y, y} f_2 h_2 u + f_2h_{2, y}f_{2, y}u
    - 2f_{2, y}^2 h_2 u + 2h_2f_2f_{2, y} u_y) z_{22}}{f_2^3u_y}\\
  &+ \frac{u_t(f_2uh_{2, y} - 2h_2f_{2, y}u + 2h_2f_2u_y) z22_y}{f_2^2 u_y}
    + \frac{h_2 uu_tz_{22,yy}}{2f_2u_y}\\
  &+ \frac{u_t (2 h_2u_y^2\ln(u_y) - 2h_2u_y^2 - 4uu_yh_{2, y}
    - h_{2, y, y} u^2)}{4u_y},
\end{align*}
while the inverse action generates the nonlocal symmetry
\begin{equation*}
  \Phi = \frac{(k_1 f_{1, x} - 2f_1k_{1, x}) u_x z_1}{f_1^2}
  - \frac{k_1u_xz_{1,x}}{f_1} + \frac{k_1}{u_x^2},
\end{equation*}
where $k_1 = k_1(x)$

\subsection{$3A = 2C$, $3B = -2C$}
\label{sec:3a-=-2c-2}

The action of the operators~$\mathcal{R}_1$, $\mathcal{R}_2$, $\mathcal{R}_3$
is as follows:
\begin{equation*}
  \xymatrix{\dots\ar@<.5ex>[r]^-{\mathcal{R}_{i}}
    &\ar@<.5ex>[r]^-{\mathcal{R}_{i}}\ar@<.5ex>[l]^-{\mathcal{R}_{i}}
    \Phi_{2i} + \Phi_{3i} + \phi_{00} + \phi_{01} + \phi_1 + \phi_2 + \phi_3
    &\ar@<.5ex>[r]^-{\mathcal{R}_{i}}\ar@<.5ex>[l]^-{\mathcal{R}_{i}}0
    &\ar[r]^-{\mathcal{R}_{i}}\ar@<.5ex>[l]^-{\mathcal{R}_{i}}\phi_2
    &0\rlap{,}
  }
\end{equation*}
where
\begin{align*}
  \Phi_{12}
  &=\frac{u_y(4f_{2, y}g_2 + 5g_{2, y}f_2) z_2}{f_2} + g_2 u_y z_{2,y}
    + \frac{2f_2 g_2}{3\cdot\sqrt{u_y}},\\
  \Phi_{13}
  &=\frac{3g_3f_{3, t}u_tz_{31}}{2f_3} - \frac{3 g_3 u_t z_{31,t}}{2}
    + \frac{3g_3u_t(4f_3 u_t\ln u_t + f_3u_t - 2f_{3, t} u)}{4}
    - \frac{3 f_3  g_{3, t}u u_t}{2};\\
  \Phi_{22}
  &= -\frac{(9f_{2, y}g_2u_y  - 5g_{2, y}f_2) z_2}{20f_2^2}
    + \frac{9g_2u_yz_{2,y}}{20f_2} + \frac{3 g_2}{10\sqrt{u_y}},\\
  \Phi_{23}
  &= -\frac{3f_{3, t} g_3u_tz_{31}}{2 f_3^2}
    + \frac{3g_3u_tz_{31,t}}{2 f_3}
    -\frac{3g_3u_t^2(4\ln u_t + 1)}{4} + \frac{3g_{3, t}u u_t}{2};\\
  \Phi_{32}
  &= \frac{u_y (4f_{2, y}g_2 + 5g_{2, y}f_2) z_2}{f_2} + g_2 u_y z_{2,y}
    + \frac{2 f_2 g_2}{3\cdot\sqrt{u_y}},\\
  \Phi_{33}
  &= \frac{3f_{3, t}g_3 u_tz_{31}}{2 f_3} - \frac{3g_3 u_t z_{31,t}}{2}
    + \frac{3g_3u_t(4f_3 u_t\ln u_t + f_3u_t - 2f_{3, t}u)}{4}
    - \frac{3 f_3 g_{3, t}u u_t}{2}.
\end{align*}
The direct action of the structure~$\mathcal{J}$ from
Section~\ref{sec:3a-=-2c} produces the nonlocal cosymmetries
\begin{align*}
  \Psi_0
  &= -\frac{k_3(2f_3 u_t - f_{3, t}u)}{f_3^2} - \frac{k_{3, t}u}{f_3},\\
  \Psi_1
  &= -\frac{f_{3, t}z_{31}}{3f_3^2} + \frac{z_{31,t}}{3 f_3}
    + \frac{2 u_t (\ln u_t + 5)}{3},\\
  \Psi_2
  &= \frac{27u_tz_1}{4 f_1 \sqrt{u_x}}
    - \frac{(2 f_3 u_t - f_{3, t}u) z_{31}}{3 f_3^2}
    - \frac{uz_{31,t}}{3 f_3}
    - \frac{4 z_{32}}{f_3}\\
  &+ \frac{u(2f_3 u_t\ln u_t + 12 f_3 u_t + 3f_{3, t} u)}{3 f_3},\\
  \Psi_3
  &= \frac{(2f_{3, t}f_3g_3u_t - 2f_{3, t}^2g_3u
    + f_{3, t}g_{3, t}f_3u + g_3f_{3, t, t}f_3u)z_{31}}{3 f_3^3}\\
  &- \frac{(2f_3g_3u_t - 2f_{3, t}g_3 u + f_3g_{3, t}u) z_{31,t}}{3f_3^2}
    - \frac{g_3 u z_{31,tt}}{3 f_3}\\
  &+ \frac{2 (g_3f_{3, t} - f_3g_{3, t}) z_{32}}{f_3^2}
    - \frac{2 g_3 z_{32,t}}{f_3}\\
  &- \frac{(2f_3^2g_{3, t, t} + 3f_{3, t}^2g_3 - 3f_{3, t}g_{3, t}f_3
    - 3 g_3f_{3, t, t}f_3) u^2}{6f_3^2}\\
  &+ \frac{(3f_3g_3u_{tt} + 3f_{3, t}g_3u_t - f_3g_{3, t}u_t)u}{3f_3}
    + \frac{g_3u_t^2(4\ln u_t - 1)}{6},
\end{align*}
while the inverse action leads to the nonlocal symmetry
\begin{equation*}
  \Phi =
  \frac{3k_2f_2(3u_y^{3/2}z_{2,y} - 3f_{2, y}u_y^{3/2}z_2 +
    2f_2^2)}{20f_2^2\sqrt{u_y}}
  + \frac{9k_{2, y}u_y z_2}{4f_2};
\end{equation*}
here~$k_2 = k_2(y)$, $k_3 = k_3(t)$.

\subsection{$3A = -C$, $3B = C$}
\label{sec:3a-=-c-2}

The recursion operators~$\mathcal{R}_{ij}$, $i = 1$, $2$, $j = 1$, $2$, $3$
from Section~\ref{sec:3a-=-2c} act as follows:
\begin{equation*}
  \xymatrix{\dots\ar@<.5ex>[r]^-{\mathcal{R}_{1i}}
    &\ar@<.5ex>[r]^-{\mathcal{R}_{1i}}\ar@<.5ex>[l]^-{\mathcal{R}_{1i}}
    \Phi_{11} + 6\Phi_{12} + \phi_1 + \phi_2
    &\ar@<.5ex>[r]^-{\mathcal{R}_{1i}}\ar@<.5ex>[l]^-{\mathcal{R}_{1i}}0
    &\ar[r]^-{\mathcal{R}_{1i}}\ar@<.5ex>[l]^-{\mathcal{R}_{i}}\phi_1
    &0\rlap{,}\\
    \dots\ar@<.5ex>[r]^-{\mathcal{R}_{2i}}
    &\ar@<.5ex>[r]^-{\mathcal{R}_{i}}\ar@<.5ex>[l]^-{\mathcal{R}_{2i}}
    -\Phi_{11} - 2\Phi_{12} + \phi_{00} + \phi_1 + \phi_3 + \phi_3
    &\ar@<.5ex>[r]^-{\mathcal{R}_{i}}\ar@<.5ex>[l]^-{\mathcal{R}_{2i}}0
    &\ar[r]^-{\mathcal{R}_{2i}}\ar@<.5ex>[l]^-{\mathcal{R}_{2i}}\phi_2
    &0\rlap{,}
  }
\end{equation*}
where
\begin{equation*}
  \Phi_{11} =\frac{k_1(12f_1u_x^3z_{1,x} + 12f_{1, x}u_x^3z_1 +
  f_1^2)}{4f_1u_x^2},\quad
  \Phi_{12} = u_xz_1k_{1, x},\quad k_1 = k_1(x).
\end{equation*}

The direct action of $\mathcal{J}$ from Section~\ref{sec:3a-=-c} gives the
nonlocal cosymmetries
\begin{align*}
  \Psi_1
  &= \frac{f_{3, t}z_{31}}{f_3^2} - \frac{z_{31,t}}{f_3}
    + \frac{(\ln u_t - 4) u_t}{3},\\
  \Psi_2
  &= -\frac{(f_{3, t} u - 2 f_3 u_t) z_{31}}{f_3^2}
    + \frac{uz_{31,t}}{f_3} + \frac{4 z_{32}}{f_3}
    - \frac{f_{3, t} u^2}{f_3} + \frac{(\ln u_t - 12)uu_t}{3},\\
  \Psi_3
  &= \frac{(2f_{3, t}^2g_3u - f_{3, t}g_{3, t}f_3 u
    - 2f_{3, t}f_3 g_3u_t - f_{3, t, t}f_3 g_3 u)z_{31}}{f_3^3}\\
  &- \frac{(2g_3f_{3, t}u - g_{3, t}f_3 u - 2 f_3 g_3u_t) z_{31,t}}{f_3^2}
    + \frac{g_3 u z_{31,tt}}{f_3}\\
  &-\frac{2(g_3f_{3, t} - g_{3, t}f_3)z_{32}}{f_3^2}
    + \frac{2 g_3 z_{32,t}}{f_3}\\
  &+ \frac{(3f_{3, t}^2 g_3 - 3f_{3, t}g_{3, t}f_3 - f_3^2g_{3, t, t}
    - 3f_{3, t, t}f_3 g_3) u^2}{6f_3^2}\\
  &- \frac{(3f_{3, t}g_3u_t + 5g_{3, t}f_3u_t + 3g_3f_3u_{tt})u}{3f_3}
    + \frac{g_3 (\ln u_t - 7) u_t^2}{3},
\end{align*}
while the inverse action produces the nonlocal symmetries~$\Phi_{11}$,
$\Phi_{12}$ and
\begin{equation*}
  \Phi_{13} =\frac{k_1(12f_1z_{1,x}u_x^3 - 12f_{1, x}u_x^3z_1 + f_1^2)}{6 u_x^2f_1^2} + \frac{4k_{1, x}u_xz_1}{f_1}.
\end{equation*}

\subsection{$A = C$, $2B = -1$}
\label{sec:a-=-c-20}

The action of the recursion operators described in Section~\ref{sec:a-=-c-6}
reads
\begin{equation*}
  \xymatrix{\dots\ar@<.5ex>[r]^-{\mathcal{R}_{1},\mathcal{R}_{3}}
    &\ar@<.5ex>[r]^-{\mathcal{R}_{1},\mathcal{R}_{3}}\ar@<.5ex>[l]^-{\mathcal{R}_{1}\mathcal{R}_{3}}
    \Phi_{20} + \Phi_3 + \phi_{00} + \phi_{01} + \phi_1 + \phi_2 + \phi_3
    &\ar@<.5ex>[r]^-{\mathcal{R}_{1},
      \mathcal{R}_{3}}\ar@<.5ex>[l]^-{\mathcal{R}_{1}, \mathcal{R}_{3}}0
    &\ar[r]^-{\mathcal{R}_{1},\mathcal{R}_{3}}\ar@<.5ex>[l]^-{\mathcal{R}_{i},
    \mathcal{R}_{3}}\phi_1
    &0\rlap{,}\\
    \dots\ar@<.5ex>[r]^-{\mathcal{R}_{2}}
    &\ar@<.5ex>[r]^-{\mathcal{R}_{2}}\ar@<.5ex>[l]^-{\mathcal{R}_{2}}
    k_0\Phi_{21} + \Phi_{22} + \Phi_3 + \phi_{00} + \phi_{01} + \phi_{01} +\phi_1
    + \phi_2 + \phi_3 
    &\ar@<.5ex>[r]^-{\mathcal{R}_{2}}\ar@<.5ex>[l]^-{\mathcal{R}_{2}}0
    &\ar[r]^-{\mathcal{R}_{2i}}\ar@<.5ex>[l]^-{\mathcal{R}_{2}}\phi_2
    &0\rlap{,}
  }
\end{equation*}
where
\begin{align*}
  \Phi_{20}
  &= \frac{k_2(f_2z_{2,y} - f_{2, y}z_2)u_y}{f_2^2}
    + \frac{2k_{2, y}u_yz_2}{f_2},\\
  \Phi_{21}
  &= \frac{f_2u_y(f_2z_{2,y} - f_{2, y}z_2)}{f_2^3},\\
  \Phi_{22}
  &= -\frac{u_y(2f_2z_{2,y} - 6f_{2, y}) k_2z_2}{f_2^3}
    + \frac{4k_{2, y}u_yz_2 }{f_2^2},\\
  \Phi_3
  &= \frac{k_3(6f_3^2u_t^{5/2} + 4f_{3, t}u_t z_3 - 5f_3u_tz_{3,t})}{5f_3}
    - \frac{k_{3, t}u_t z_3}{5};
\end{align*}
here $k_0 = \const$, $k_2 = k_2(y)$, $k_3 = k_3(t)$.

The direct action of the structure~$\mathcal{J}$ gives the nonlocal symmetry
\begin{equation*}
  \Phi_2 = \frac{2u_yz_{22}}{5g_2^{3/5}} + \frac{g_2^{2/5}}{5 u_y^2}.
\end{equation*}
The inverse action generates three nonlocal cosymmetries
\begin{align*}
  \Psi_1
  &= -\frac{3u_tz_1}{2f_1\sqrt{u_x}} + \frac{3\sqrt{u_t}z_3}{2f_3}
    - 2 u_t u,\\
  \Psi_2
  &= \frac{(2g_1f_{1, x} - f_1g_{1, x})u_tz_1}{2f_1^2\sqrt{u_x}}
    - \frac{g_1u_tz_{1,x}}{f_1\sqrt{u_x}},\\
  \Psi_3
  &= -\frac{(2g_3f_{3, t} - f_3g_{3, t})\sqrt{u_t} z_3}{2f_3^2}
    + \frac{g3 z_{3,t}\sqrt{u_t}}{f_3}.
\end{align*}

\subsection{$A = -2C$, $B = -2C$}
\label{sec:a-=-2c-2}

The action of the recursion operators~$\mathcal{R}_i$, $i = 1$, $2$, $3$, from
Section~\ref{sec:a-=-2c} is presented by the diagram
\begin{equation*}
  \xymatrix{\dots\ar@<.5ex>[r]^-{\mathcal{R}_{i}}
    &\ar@<.5ex>[r]^-{\mathcal{R}_{i}}\ar@<.5ex>[l]^-{\mathcal{R}_{i}}
    \Phi_{2} + \Phi_3 + \phi_{00} + \phi_{01} + \phi_1 + \phi_2 + \phi_3
    &\ar@<.5ex>[r]^-{\mathcal{R}_i}\ar@<.5ex>[l]^-{\mathcal{R}_{i}}0
    &\ar[r]^-{\mathcal{R}_{i}}\ar@<.5ex>[l]^-{\mathcal{R}_{i}}\phi_1
    &0\rlap{,}
  }
\end{equation*}
where
\begin{align*}
  \Phi_2
  &= \frac{k_2(6f_2^2u_y^{5/2}  + 5f_2u_yz_{2,y} - 5f_{2, y}u_yz_2)}{4f_2^2}
    + \frac{k_{2, y}u_yz_2}{4f_2},\\
  \Phi_3
  &= \frac{k_3(f_3z_{3,t} +  f_{3, t}z_3)u_t}{2f_3} + k_{3, t}u_tz_3.
\end{align*}
The structure~$\mathcal{J}$ gives the symmetry
\begin{equation*}
  \Phi_3' = \frac{k_3(f_3z_{3,t} - f_{3, t}z_3)u_t}{f_3^2}
  + \frac{2k_{3, t}u_tz_3}{f_3}
\end{equation*}
by the direct action. The inverse action produces the nonlocal cosymmetries
\begin{align*}
  \Psi_1
  &= \frac{3u_tz_1}{2f_1\sqrt{u_x}} + \frac{3u_tz_2}{2f_2\sqrt{u_y}} + 2uu_t,\\
  \Psi_2
  &= \frac{u_t(-h_{1, x}f_1 + 2f_{1, x}h_1) z_1}{2f_1^2\sqrt{u_x}}
    - \frac{h_1u_tz_{1,x}}{f_1\sqrt{u_x}},\\
  \Psi_3
  &= \frac{u_t(2f_{2, y}h_2 - h_{2, y}f_2)z_2}{2f_2^2\sqrt{u_y}}
    - \frac{h_2u_tz_{2,y}}{f_2\sqrt{u_y}},
\end{align*}
where $h_1 = h_1(x)$, $h_2 = h_2(y)$.

\begin{remark}
  The structures presented in Sections~\ref{sec:a-=-c-6} and~\ref{sec:a-=-2c}
  are invertible. The inverse ones are
  \begin{gather*}
    \mathcal{J}_1= u_t\phi - \frac{u_tw_{11}}{f_1\sqrt{u_x}}
    - \frac{\sqrt{u_t}w_{13}}{f_3}
  \end{gather*}
  and
  \begin{gather*}
    \mathcal{J}_2 =  u_t \phi - \frac{u_tw_{21}}{2 f_1 \sqrt{u_x}}
    - \frac{u_t w_{22}}{2 f_2\sqrt{u_y}},
  \end{gather*}
  respectively. The nonlocal variables~$w_{ij}$ are defined by
  \begin{align*}
    w_{11,y}
    &= -\frac{f_1(2u_xu_{yt} - u_yu_{xt})q}{4 u_t\sqrt{u_x}}
      - \frac{f_1 u_yq_x}{2\sqrt{u_x}},\\
      w_{11,t}
    &= \frac{f_1 u_{xt}q}{2\sqrt{u_x}}
      - \frac{f_1 u_tq_x}{\sqrt{u_x}} - f_1\sqrt{u_x}q_t,\\
    w_{31,x}
    &= \frac{f_3 u_{xt}q}{\sqrt{u_t}}
      - \frac{f_3u_xq_t}{\sqrt{u_t}},\\
    w_{31,y}
    &= \frac{f_3 u_{yt}q}{\sqrt{u_t}}
      + f_3\sqrt{u_t}q_y - \frac{f_3u_yq_t}{2\sqrt{u_t}},\\
    w_{12,y}
    &= \frac{2f_1 (2 u_x u_{yt} + 2 u_yu_{xt})q}{u_t\sqrt{u_x}}
      - \frac{2 f_1 u_yq_x}{\sqrt{u_x}},\\
    w_{12,t}
    &= \frac{2 f_1 u_{xt}q}{\sqrt{u_x}}
      - \frac{f_1 u_tq_x}{\sqrt{u_x}} + 2 f_1\sqrt{u_x}q_t,\\
    w_{22,x}
    &= \frac{2f_2 (u_xu_{yt} + u_yu_{xt})q}{u_t\sqrt{u_y}}
      - 2 f_2\sqrt{u_y}q_x
      - \frac{2 f_2u_xq_y}{\sqrt{u_y}},\\
    w_{22,t}
    &= \frac{f_2 u_{yt}q}{\sqrt{u_y}} -\frac{f_2u_tq_y}{\sqrt{u_y}}.
  \end{align*}
\end{remark}

%

\section*{Conclusion}
\label{sec:conclusion}

Let us briefly summarize the results obtained.
\begin{center}
  \begin{tabular}{|c|c|c|c|}
    \hline
    The case
    &$\mathcal{T}\leftrightarrow\mathcal{T}$
    &$\mathcal{T}^*\leftrightarrow\mathcal{T}^*$
    &$\mathcal{T}^*\leftrightarrow\mathcal{T}$\\
    \hline
    $A + B + C = 0$
    &3&3&9\\
    \hline
    $A = -C,\quad  B = -C$
    &3&--&1\\
    \hline
    $A = -C,\quad  B = C$
    &1&--&1\\
    \hline
    $A = -C,\quad  2B = -3C$
    &2&--&1\\
    \hline
    $A = -C,\quad  B = -3C$
    &5&--&1\\
    \hline
    $A = -C,\quad  2B = 3C$
    &3&--&1\\
    \hline
    $A = -C,\quad  B = 3C$
    &6&--&1\\
    \hline
    $3A = 2C,\quad  3B = -2C$
    &3&--&1\\
    \hline
    $3A = -C,\quad  3B = C$
    &6&--&1\\
    \hline
    $A = C,\quad  2B = -C$
    &3&--&2\\
    \hline
    $A = -2C,\quad  B = -2C$
    &3&--&2\\
    \hline
  \end{tabular}
\end{center}
This table presents the number of integrable structures found in each case.

\section*{Acknowledgements}
\label{sec:acknowledgements}

Computations were done with the \textsc{Jets} software,~\cite{BM-Jets}.

\appendix

\section{Nonlocal variables in extended coverings}
\label{sec:nonl-vari-extend}

We present here defining equations for the nonlocal variables that arise in
the extensions. We also provide some integrable structures whos expressions
are too long for the main text.

\subsection{$A + B + C = 0$}\label{sec:a-+-b-1}

The other six structures are
\begin{align*}
  \mathcal{J}_{12}
  &= \frac{af_1(au_xu_{xt} - u_tu_{xx})}{2u_x^4}q + \frac{af_1u_t}{2
    u_x^3}q_x - \frac{a^2f_1}{2u_x^2}q_t \\
  &- \frac{af_1u_t(u_tu_{xx} - u_xu_{xt})}{2f_3u_x^4}w3
    -\frac{f_{1, x}u_t}{f_1u_x^2}w13 + \frac{u_t}{u_x^2}w_{13,x},\\
  \mathcal{J}_{13}
  &= \lambda_{13}q
    + \frac{a(af_1u_xu_{xt} + 2f_{1, x}u_tu_x -
    2f_1u_tu_{xx})}{2u_x^4)}q_x \\
  &- \frac{a^2(2f_{1, x}u_x - f_1u_{xx})q_t}{2u_x^3}
    + \frac{af_1u_t}{2u_x^3}q_{xx} - \frac{a^2f_1}{2u_x^2}q_{xt}\\
  &-\frac{f_{1, x, x}u_t}{f_1u_x^2}w13 -
    \frac{\mu_{13}}{f_3} w3 + \frac{u_t}{u_x^2} w_{13,xx},\\
  \intertext{where}
  \lambda_{13}
  &= \frac{a (2af_{1, x}u_{xt} + f_1u_{xxt})}{2u_x^3}
    - \frac{f_1(2au_{xt}u_{xx} +  u_tu_{xxx}) + 2f_{1, x}u_tu_{xx}}{2u_x^4} +
    \frac{f_1u_tu_{xx}^2}{u_x^5},\\
  \mu_{13}
  &= -\frac{au_t(2f_{1, x}u_{xt}  + f_1u_{xxt})}{2u_x^3}
    + \frac{f_1(u_tu_{xxx} + 2u_{xt}u_{xx}) + 2f_{1,x}u_tu_{xx}}{2u_x^4} -
    \frac{f_1u_tu_{xx}^2}{u_x^5},\\
  \mathcal{J}_{22}
  &= (a - 1)\left(\frac{(a - 1)^2f_2u_{xt}}{2u_y^2u_x}
    - \frac{a(a - 1)f_2u_{yt}}{2u_y^3} + \frac{af_2u_tu_{yy}}{2u_y^4}
    - \frac{af_{2, y}u_t}{2u_y^3}\right) q\\
  &+ \frac{(a - 1)^2f_2u_t}{2u_y^2u_x}q_x -
    \frac{(a - 1)af_2u_t}{2u_y^3}q_y- \lambda_{22}w_1 +
    \frac{u_t}{u_y^2}w_{21,y}\\
  \intertext{where}
  \lambda_{22}
  &= -\frac{(a - 1)^2f_2u_y^2u_{xt}}{f_1u_y^4}
    + \frac{af_2u_xu_{yt}}{f_1u_y^3} -
    \frac{f_2u_tu_xu_{yy}}{f_1u_y^4}
    + \frac{f_{2,y} u_t u_x}{f_1u_y^3},\\
  \mathcal{J}_{23}
  &= \lambda_{23} q
    + \frac{f_2(au_yu_{yt} - u_tu_{yy}) + 2f_{2, y}u_yu_t}{af_2u_y^3u_x}q_x\\
  &- \frac{af_2u_yu_{yt} + 2f_{2, y}u_yu_t - 2f_2u_tu_{yy} - f_2u_yu_{yt}}{(a
    - 1)f_2u_y^4}q_y-
    \frac{au_xu_{yt} - (a - 1)u_yu_{xt}}{au_y^2u_xu_t}q_t\\
  &
    - \frac{a - 1}{au_yu_x}q_{xt} - \frac{u_t}{(a - 1)u_y^3}q_{yy}
    + \frac{1}{u_y^2}q_{yt} + \mu_{23} w_1 +
    \frac{2 u_t}{a(a - 1)^2f_2u_y^2} w_{21,yy},\\
  \intertext{where}
  \lambda_{23}
  &= - \frac{(a - 1)^2}{au_tu_yu_x}u_{xtt} + \frac{(a - 1)}{u_y^2u_t}u_{ytt} +
    \frac{u_tu_{yyy}}{(a - 1)u_y^4} - \frac{u_{yyt}}{u_y^3}\\
  &- \frac{(a - 1)u_{tt}(au_xu_{yt} - (a - 1)u_yu_{xt})}{au_y^2u_t^2u_x}
    + \frac{(a - 1)(af_2u_{yt} + 2f_{2, y}u_t) u_xt}{af_2u_tu_y^2u_x}\\
  &- \frac{(a - 1)u_{yt}^2}{u_tu_y^3}
    - \frac{2f_{2, y}u_{yt}}{f_2u_y^3}
    - \frac{(a - 1)u_{yy} u_{xt}}{au_y^3u_x}
    - \frac{f_{2, y, y}u_t}{(a - 1)f_2u_y^3}\\
  &+ \frac{2 u_yy (af_2u_{yt} + f_{2, y}u_t - f_2u_{yt})}{(a -1)f_2u_y^4} -
    \frac{2u_tu_{yy}^2}{(a - 1)u_y^5},\\
  \mu_{23}
  &= -\frac{2(a - 1)u_{xtt}}{af_1u_tu_y}
    + \frac{2u_tu_xu_{yyy}}{(a - 1)af_1u_y^4}
    +\frac{2u_xu_{ytt}}{f_1u_tu_y^2}
    - \frac{2u_xu_{yyt}}{(a - 1)f_1u_y^3} -
    \frac{4u_tu_xu_{yy}^2}{(a - 1)af_1u_y^5}\\
  &+ \frac{2(2af_2u_xu_{yt} - af_2u_yu_{xt} + 2f_{2, y}u_tu_x
    + f_2u_yu_{xt})u_{yy}}{af_1f_2(a - 1)u_y^4}\\
  &+ \frac{2(af_2u_tu_yt + af_2u_yu_{tt} + 2f_{2, y}u_t^2 - f_2u_yu_{tt})u_{xt}}{af_1f_2u_y^2u_t^2} -
    \frac{2u_xu_{yt}u_{tt}}{f_1u_y^2u_t^2}\\
  &- \frac{2u_xu_{yt}^2}{f_1u_tu_y^3} -
    \frac{4f_{2, y}u_xu_{yt}}{(a - 1)f_1f_2u_y^3}
    - \frac{2f_{2, y, y} u_t u_x}{(a - 1)af_1f_2u_y^3},\\
  \mathcal{J}_{32}
  &= -\frac{f_3(-au_yu_{tt} - u_tu_{yt} + u_yu_{tt}) + (a-1)u_tu_yf_{3,
    t}}{u_t^3u_y}q - \frac{f_3}{u_tu_y)}q_y - \frac{(a -1)f_3}{u_t^2}q_t\\
  &+ \frac{f_3(u_tu_{yt} - u_yu_{tt}) + f_{3, t}u_yu_t}{f_2u_t^3}w_2
    + \frac{1}{u_t}w_{32,t},\\
  \mathcal{J}_{33}
  &= \lambda_{33} q - \frac{2f_{3, t}u_t - f_3u_{tt}}{u_t^2u_y}q_y\\
  &- \frac{f_3(-2au_yu_{tt} - u_tu_{yt} + 2u_yu_{tt}) + 2(a-1)u_tu_yf_{3,
    t}}{u_t^3u_y}q_t 
    - \frac{f_3}{u_tu_y}q_{yt}\\
  &- \frac{(a -1)f_3}{u_t^2}q_{tt} - \mu_{33}w_2 + \frac{1}{u_t}w_{32,tt},\\
  \intertext{where}
  \lambda_{33}
  &= \frac{f_3u_{ytt}}{u_t^2u_y} + \frac{(a - 1)f_3u_{ttt}}{u_t^3}
    - \frac{2(a - 1)f_3u_{tt}^2}{u_t^4}
    + \frac{2((a - 1)f_{3, t}u_y - f_3u_{yt})u_{tt}}{u_t^3u_y}\\
  &+ \frac{2f_{3, t}u_{yt}}{u_t^2u_y} - \frac{(a -1)f_{3, t, t}}{u_t^2},\\
  \mu_{33}
  &= -\frac{f_3u_{ytt}}{f_2u_t^2} + \frac{f_3u_y u_{ttt}}{f_2u_t^3}
    - \frac{2f_3u_yu_{tt}^2}{f_2u_t^4}
    + \frac{2(f_{3, t}u_y + f_3u_{yt}) u_tt}{f_2u_t^3} \\
  &- \frac{2f_{3, t}u_{yt}}{f_2u_t^2} - \frac{f_{3, t, t}u_y}{f_2u_t^2}.
\end{align*}
The defining equations for~$w_{ij}$ are
\begin{align*}
  w_{11,y}
  &= \frac{(a - 1)(af_1u_x(u_xu_{yt}  - u_yu_{xt}) + f_{1, x}u_tu_yu_x -
    f_1u_tu_yu_{xx} + f_1u_yu_xu_{xt})}{2f_1u_x^2u_t}w_1\\
  &- \frac{(a - 1)u_y}{2u_x}w_{1,x}\\
  w_{11,t}
  &= -\frac{f_{1, x}u_tu_x - f_1u_tu_{xx} + f_1u_xu_{xt}}{2f_1u_x^2}w_1
    + \frac{u_t}{2u_x}w_{1,x} - \frac{(a - 1)af_1u_{xt}}{4u_x^2}q\\
  &- \frac{af_1u_t}{4u_x^2}q_x + \frac{a^2f_1}{4u_x}q_t,\\
  w_{13,y}
  &= \frac{a(a - 1)f_1(u_tu_xu_{yt} - u_tu_yu_{xt})}{2f_3u_tu_x^2}w3
    - \frac{a(a - 1)f_1u_yu_{xt}}{2u_x^2u_t}q\\
  &+ \frac{a(a - 1)f_1 u_y}{2u_tu_x}q_t,\\
  w_{13,t}
  &= \frac{af_1(f_3u_tu_{xt} + f_{3, t}u_tu_x -
    f_3u_xu_{tt})}{2f_3^2u_x^2}w_3\\
  &- \frac{af_1u_t}{2f_3u_x}w_{3,t}
    + \frac{af_1u_{xt}}{2u_x^2}q - \frac{af_1}{2u_x}q_t,\\
  w_{21,x}
  &= \nu_{21}w_1 + \frac{(a - 1)f_2u_x}{f_1u_y}w_{1,x}
    - \frac{(a - 1)af_2(au_xu_{yt} - (a - 1)u_yu_{xt})}{2u_y^2u_t} q\\
  &+ \frac{(a - 1)af_2}{2u_y}q_x,\\
  \intertext{where}
  \nu_{21}
  &= -\frac{(a - 1)f_2(af_1u_x(u_xu_{yt} - u_yu_{xt})
    + u_t u_y u_x f_{1, x} + f_1u_y(u_xu_{xt} -
    u_tu_{xx}))}{f_1^2u_y^2u_t},\\
  w_{21,t}
  &= -\frac{(a - 1)f_2(u_xu_{yt} - u_yu_{xt})}{f_1u_y^2}w_1
    - \frac{(a - 1)f_2(au_xu_{yt} - (a - 1)u_yu_{xt})}{2u_xu_y^2}q\\
  &+ \frac{(a - 1)f_2u_t}{2u_yu_x}q_x,\\
  w_{22,x}
  &= \frac{-af_2u_xu_tu_{yy} - a(a - 1)f_2u_y^2u_{xt}
    + a^2f_2u_xu_yu_yt + af_{2, y}u_tu_xu_y}{f_2u_y^2u_t}w_2-
    \frac{au_x}{u_y}w_{2,y}\\ 
  &    + \frac{(a-1)f_2^2(au_xu_{yt}-(a - 1)u_yu_{xt})}{f_2u_y^2u_t}q
    + \frac{(a - 1)^2f_2}{u_y}q_x -\frac{a(a - 1)f_2u_x}{u_y^2}q_y,\\
  w_{22,t}
  &= \frac{f_{2, y}u_yu_t - f_2u_tu_{yy} + f_2u_yu_{yt}}{f_2u_y^2}w_2
    - \frac{u_t}{u_y}w_{2,y},\\
  w_{32,x}
  &= -\frac{af_3(u_xu_{yt} - u_yu_{xt})}{f_2u_t^2}w_2
    - \frac{af_3u_xu_{yt}}{u_yu_t^2}q + \frac{af_3u_x}{u_tu_y}q_y,\\
  w_{32,y}
  &= \frac{f_3(f_{2, y} u_y u_t - f_2u_tu_{yy} + f_2u_yu_{yt})}{f_2^2u_t^2}w_2
    - \frac{f_3u_y}{f_2u_t}w_{2,y}
    - \frac{(a - 1)f_3u_{yt}}{u_t^2}q + \frac{(a - 1)f_3}{u_t}q_y,\\
  w_{33,x}
  &= \frac{a(f_3u_tu_{xt} + f_{3, t}u_tu_x - f_3u_xu_{tt})}{f_3u_t^2}w_3
    - \frac{au_x}{u_t}w_{3,t},\\
  w_{33,y}
  &= \frac{(a - 1)(f_{3, t}u_yu_t + f_3u_tu_{yt} - f_3u_yu_{tt})}{/(u_t^2 f_3)}w_3
    - \frac{(a - 1)u_y}{u_t}w_{3,t}\\
  &- \frac{af_3u_{yt}}{u_t^2}q + \frac{f_3}{u_t}q_y
    + \frac{(a - 1)f_3u_y}{u_t^2}q_t.
\end{align*}

\subsection{$A = -C$, $B = -C$}\label{sec:a-=-c-7}

The rest two recursion operators are
\begin{align*}
  \mathcal{R}_2
  &= \left(4f_3u_t^2u_{tt} + 2f_{3, t}u_t^3 -\frac{2f_3u_t^3u_{xt}}{u_x}
    - \frac{f_3u_t^3u_{yt}}{u_y}\right)q + \frac{2f_3u_t^4}{u_x}q_x +
    \frac{f_3u_t^4}{u_y}q_y\\ 
  & + \rho_{21}w_{11} + \rho_{22}w_{11,x}
    + \left(\frac{2f_3u_t^3u_{xt}}{f_1u_x^2} - \frac{4f_3u_t^2u_{tt}}{f_1u_x}
    - \frac{2f_{3, t}}{f_1u_x}u_t^3\right)w_{12}\\
  & - \left(\frac{2f_3u_t^2u_{tt}}{f_2} + \frac{u_t^3f_{3,
    t}}{f_2}\right)w_{21}
    + u_tw_{31,t},
    \intertext{where}
    \rho_{21}
  &= -\frac{f_{1, x}f_3u_t^3uu_{xt}}{f_1^2u_x^2}
    - \frac{2f_3u_t^2(f_1u_x - f_{1, x}u)u_{tt}}{f_1^2u_x}
    + \frac{f_3f_{1, x}u_t^4}{f_1^2 u_x}\\
  &- \frac{f_{3, t}(f_1u_x - f_{1, x}u) u_t^3}{f_1^2u_x},\\
  \rho_{22}
  &= -\frac{2f_3uu_t^2u_{tt}}{f_1u_x}
    + \frac{f_3uu_t^3u_{xt}}{f_1u_x^2}
    - \frac{f_3u_t^4}{f_1u_x} - \frac{f_{3, t} uu_t^3}{f_1u_x},
\end{align*}
and
\begin{align*}
  \mathcal{R}_3
  &= \rho_{31}q
    - \left(\frac{f_3u_t^5u_{xx}}{u_x^3} - \frac{10f_3u_t^3u_{tt}}{u_x}
    + \frac{3f_3u_t^4u_{xt}}{u_x^2} - \frac{4f_{3, t}u_t^4}{u_x}\right)q_x \\
  &+\frac{u_t^3(2f_{3, t}u_tu_y  - f_3u_tu_{yt} + 5f_3u_yu_{tt})}{u_y^2}q_y
    - \frac{f_3u_t^3(u_xu_{yt} + u_yu_{xt})}{u_xu_y)}q_t
    + \frac{f_3u_t^5}{u_x^2}q_{xx}\\
  &+ \frac{f_3u_t^4}{u_x}q_{xt} + \frac{f_3u_t^4}{u_y}q_{yt} + \rho_{32}w_{11}
    - u_t\rho_{33}w_{11,x} - \rho_{34}w_{12}\\
  &- \left(\frac{2f_3u_t^2u_{ttt}}{f_2} + \frac{2f_3u_tu_{tt}^2}{f_2}
    + \frac{4f_{3, t}u_t^2u_{tt}}{f_2}
    + \frac{u_t^3f_{3, t, t}}{f_2}\right)w_{21} + u_t w_{31,tt}
    \intertext{where}
    \rho_{31}
  &= -\frac{f_3u_t^4u_{xxt}}{u_x^2} - \frac{2f_3u_t^3u_{xtt}}{u_x}
    - \frac{f_3u_t^3u_{ytt}}{u_y} + 4f_3u_t^2u_{ttt}
    + \frac{f_3u_t^3u_{yt}^2}{u_y^2}\\
  &- \frac{2u_t^2(f_{3, t}u_t + 2f_3u_{tt})u_{yt}}{u_y}
    + \frac{f_3u_t^4u_{xt}u_{xx}}{u_x^3} + 4f_3u_tu_{tt}^2\\
  &+ \frac{8u_t^2(f_{3, t}u_x - f_3u_{xt})u_{tt}}{u_x}
    + \frac{3f_3u_t^3u_{xt}^2}{u_x^2} -
    \frac{4f_{3, t}u_t^3u_{xt}}{u_x} + 2f_{3, t, t}u_t^3,\\
  \rho_{32}
  &= -\frac{2u_t^2f_3(f_1u_x - f_{1, x}u)u_{ttt}}{f_1^2u_x}
    - \frac{f_{1, x}f_3uu_t^3u_{xtt}}{f_1^2u_x^2}
    + \frac{2f_{1, x}f_3uu_t^3u_{xt}^2}{f_1^2u_x^3}\\
  &- \frac{2u_t^2f_{1, x}(f_3u_t^2 + f_{3, t}uu_t +
    2f_3uu_{tt}) u_{xt}}{f_1^2u_x^2}
    + \frac{2f_3u_t(f_{1, x}u - f_1u_x)u_{tt}^2}{f_1^2u_x}\\  
  &+ \frac{u_t^2(5f_3f_{1, x} u_t - 4f_1f_{3, t}u_x + 4f_{1, x}f_{3,
    t}u)u_{tt}}{f_1^2u_x}\\ 
  &+
    \frac{u_t^3(2f_{3, t}f_{1, x}u_t - f_1f_{3, t, t}u_x
    + f_{3, t, t}f_{1, x}u)}{f_1^2u_x},\\
  \rho_{33}
  &= -\frac{f_3uu_t^2u_{xtt}}{f_1u_x^2} + \frac{2f_3uu_tu_{ttt}}{f_1u_x} +
    \frac{2f_3uu_t^2u_{xt}^2}{f_1u_x^3} -\frac{4f_3uu_tu_{tt}}{f_1u_x^2}\\
  &-
    \frac{2u_t^2(f_{3, t}u + f_3u_t) u_xt}{f_1u_x^2}
    + \frac{2f_3uu_{tt}^2}{f_1u_x}
    + \frac{(5f_3u_t + 4f_{3, t}u)u_tu_{tt}}{f_1u_x}\\
  &+ \frac{u_t^2(2f_{3, t}u_t + f_{3, t, t}u)}{f_1u_x},\\
  \rho_{34}
  &= -\frac{2f_3u_t^3u_{xtt}}{f_1u_x^2}
    + \frac{4f_3u_t^2u_{ttt}}{f_1u_x}
    + \frac{4f_3u_t^3u_{xt}^2}{f_1u_x^3}
    - \frac{4u_t^2(f_{3, t}u_t + 2f_3u_{tt})u_{xt}}{f_1u_x^2}\\
  &+ \frac{4f_3u_tu_{tt}^2}{f_1u_x}
    + \frac{8f_{3, t}u_t^2u_tt}{f_1u_x}
    + \frac{2f_{3, t, t}u_t^3}{f_1u_x}.
\end{align*}
The defining equations for~$w_{ij}$ are
\begin{align*}
  w_{11,y}
  &= \frac{f_1(u_xu_{yt} + u_yu_{xt})}{u_xu_t}q - \frac{f_1u_y}{u_x}q_x,\\
  w_{11,t}
  &= \frac{f_1u_{xt}}{u_x}q - \frac{f_1u_t}{u_x}q_x + f_1q_t,\\
  w_{12,y}
  &= f_1\nu_{21} q - \frac{f_1u_y(uu_{xx} + 2u_x^2)}{2u_x^2}q_x
    + \frac{f_1uu_y}{2u_x}q_{xx},\\
  \intertext{where}
  \nu_{21}
  &= -\frac{uu_yu_{xxt}}{2u_tu_x} - \frac{uu_yu_{xtt}}{2u_t^2}
    - \frac{uu_xu_{ytt}}{2u_t^2} + \frac{u(u_xu_{yt} +
    u_yu_{xt})u_{tt}}{2u_t^3} \\
  &+ \frac{uu_yu_{xt}u_{xx}}{2u_tu_x^2} - \frac{(uu_{yt} - u_tu_y)u_{xt}}{u_t^2}
    + \frac{u_xu_{yt}}{u_t},\\ 
  w_{12,t}
  &= -\frac{f_1(uu_xu_{xxt} - uu_{xt}u_{xx} - u_x^2u_{xt})}{2u_x^2}q
    - \frac{f_1u_t(uu_{xx} + u_x^2)}{2u_x^2}q_x\\
  &+ \frac{f_1u_x}{2}q_t + \frac{f_1uu_t}{2u_x}q_{xx} - \frac{f_1u}{2}q_{xt},\\
  w_{21,x}
  &= \frac{f_2(u_xu_{yt} + u_yu_{xt})}{u_tu_y}q - f_2q_x -
    \frac{f_2u_x}{u_y}q_y,\\ 
  w_{21,t}
  &= \frac{f_2u_{yt}}{u_y}q - \frac{f_2u_t}{u_y}q_y,\\
  w_{22,x}
  &= \nu_{22}q - (f_{2, y}u + 2f_2u_y)q_x - \frac{f_2u_x(uu_{yy} + 2 u_y^2)}{u_y^2}
    q_y + \frac{f_2uu_x}{u_y}q_{yy},\\ 
  \intertext{where}
  \nu_{22}
  &= -\frac{f_2uu_yu_{xtt}}{u_t^2} - \frac{f_2uu_xu_{yyt}}{u_tu_y}
    - \frac{f_2uu_xu_{ytt}}{u_t^2}
    - \frac{f_2u(2u_tu_{yt} - u_yu_{tt})u_{xt}}{u_t^3}\\
  &+ \frac{f_2uu_xu_{yt}u_{tt}}{u_t^3}
    + \frac{f_2uu_xu_{yt}u_{yy}}{u_tu_y^2} - f_{2, y}u_x,\\
  w_{22,t}
  &= -\frac{f_{2, y}u_tu_y^2 + f_2uu_yu_{yyt} - f_2uu_{yt}u_{yy} +
    f_2u_y^2u_{yt}}{u_y^2}q\\
  &- \frac{f_2u_t(uu_{yy} + u_y^2)}{u_y^2}q_y
    - (f_{2, y}u + f_2u_y)q_t + \frac{f_2uu_t}{u_y}q_{yy} - f_2uq_{yt},\\
  w_{31,x}
  &= \frac{f_3u_t(u_xu_{yt} - 3u_yu_{xt})}{u_y}q
    - 3f_3u_t^2q_x - \frac{f_3u_t^2u_x}{u_y}q_y + f_3u_t \nu_{31}w_{11}\\
  &- \frac{f_3u_t(2f_{1, x}uu_tu_x + f_1uu_tu_{xx} - 2f_1uu_xu_{xt}
    - 2f_1u_tu_x^2)}{f_1^2u_x^2)}w_{11,x}\\
  &- \frac{2f_3u_t(f_{1, x}u_tu_x + f_1u_tu_{xx} -
    2f_1u_xu_{xt})}{f_1^2u_x^2}w_{12}\\
  &+ \frac{f_3uu_t^2}{f_1u_x}w_{11,xx}
    + \frac{2f_3u_t^2}{f_1u_x}w_{12,x} + \frac{2f_3u_tu_{xt}}{f_2}w_{21},\\
  \intertext{where}
  \nu_{31}
  &= \frac{f_{1, x} uu_tu_{xx}}{f_1^2u_x^2}
    - \frac{2(f_{1, x}u - f_1u_x)u_{xt}}{f_1^2u_x}\\
  &+ \frac{(2f_{1, x}^2 u - 2f_{1, x}f_1u_x - f_{1, x, x}f_1u)u_t}{f_1^3u_x},\\
  w_{31,y}
  &= -\frac{f_3u_t(u_xu_{yt} - 3u_yu_{xt})}{u_x}q
    - \frac{3f_3u_t^2u_y}{u_x}q_x - 2f_3u_t^2q_y- f_3 u_t\mu_{31}w_{11} \\
  &+ \frac{f_3u_t(uu_xu_{yt} - uu_yu_xt + u_tu_xu_y)}{f_1u_x^2}w_{11,x}
    - \frac{f_3u_t(f_{2, y}u_t - 2f_2u_{yt})}{f_2^2}w_{21}\\
  &+ \frac{2f_3u_t(u_xu_{yt} - u_yu_{xt})}{f_1u_x^2}w_{12}
    + \frac{f_3u_t^2}{f_2}w_{21,y},
    \intertext{where}
    \mu_{31}
  &= \frac{(f_{1, x}u - 2f_1u_x)u_{yt}}{f_1^2u_x}
    - \frac{f_{1, x}uu_yu_{xt}}{f_1^2u_x^2}
    + \frac{f_{1, x}u_tu_y}{f_1^2u_x}.
\end{align*}
The variables~$w_i$ were defined in~\cite[Section 3.2]{ABC-1}.

\subsection{$A = -C$, $B = C$}\label{sec:a-=-c-8}

These are the defining equations
\begin{align*}
  w_{11,y}
  &= \nu_{11,1}q + \nu_{11,2} q_y + \frac{f_1u_yu_x^2}{u_t}q_t
    + \frac{f_1uu_x^2}{u_y}q_{yy} - \nu_{11,3}w21 + \nu_{11,4}w_{21,y}\\
  & +
    \frac{f_1u_x(2f_2u_y^2u_{xt} + (f_{2, y}u_tu_x - 2f_2u_xu_{yt})u_y
    + f_2u_tu_xu_{yy})}{f_2^2u_y^2u_t)}w_{22}\\
  &- \frac{f_1uu_x^2}{f_2u_y} w_{21,yy}
    - \frac{2f_1u_x(u_xu_{yt} - u_yu_{xt})}{f_3u_t}w_{31}
    - \frac{f_1u_x^2}{f_2u_y} w_{22,y},
    \intertext{where}
  \nu_{11,1}
  &= -\frac{2f_1u_x(f_{2, y}u + 4f_2u_y) u_xt}{f_2u_t}
    - \frac{f_1f_{2, y}uu_x^2u_{yy}}{f_2u_y^2}\\
  &+ \frac{f_1u_x^2(2f_{2, y}u + 7f_2u_y)u_{yt}}{f_2u_tu_y}
    + \frac{f_1f_{2, y}u_x^2}{f_2}
    - \frac{f_1uu_x^2(f_{2, y}^2 - f_{2, y, y}f_2)}{f_2^2u_y},\\
  \nu_{11,2}
  &= -\frac{2f_1uu_xu_{xt}}{u_t}
    - \frac{f_1uu_x^2u_{yy}}{u_y^2}
    + \frac{2f_1uu_x^2u_{yt}}{u_yu_t}
    + \frac{f_1u_x^2(f_{2, y}u + 5f_2u_y)}{f_2u_y},\\
  \nu_{11,3}
  &= \frac{2f_1u_x(f_{2, y}u - f_2u_y)u_{xt}}{f_2^2u_t}
    + \frac{f_1f_{2, y}uu_x^2u_{yy}}{f_2^2u_y^2}
    - \frac{2f_1u_x^2(f_{2, y}u - f_2u_y)u_{yt}}{f_2^2u_yu_t}\\
  &+ \frac{f_1u_x^2(2f_{2, y}^2u - 2f_{2, y}f_2u_y - f_{2, y, y}f_2u)}{f_2^3u_y},\\
  \nu_{11,4}
  &= \frac{2f_1uu_xu_{xt}}{f_2u_t}
    + \frac{f_1uu_x^2u_{yy}}{f_2u_y^2}
    + \frac{2 f_1 u_x^2 (f_{2, y}u - f_2u_y)}{f_2^2u_y}
    - \frac{2f_1uu_x^2u_{yt}}{f_2u_yu_t},\\
  w_{11,t}
  &= -\frac{f_1u_x(u_xu_{yt} - 2u_yu_{xt})(f_{2, y}u + 4f_2u_y)}{f_2u_y^2}q
    + 3f_1u_x^2q_t\\
  &- \frac{f_1u_x((-2uu_{xt} - 3u_tu_x)u_y + uu_xu_{yt})}{u_y^2}q_y\\
  &- \frac{f_1u_x(2f_2u_y^2u_{xt} - f_{2, y}(2uu_{xt} + u_tu_x)u_y
    + f_{2, y}uu_xu_{yt})}{f_2^2u_y^2}w_{21}\\
  &+ \frac{f_1u_x(uu_xu_{yt} - 2uu_yu_{xt} - u_tu_xu_y)}{f_2u_y^2}w_{21,y}
    + \frac{f_1u_x(u_xu_{yt} - 2u_yu_{xt})}{f_2u_y^2}w_{22}\\
  &+ \frac{f_1u_x(f_{3, t}u_x - 2f_3u_{xt})}{f_3^2}w_{31} -
    \frac{f_1u_x^2}{f_3}w_{31,t},\\
  w_{21,x}
  &= \frac{f_2(u_xu_{yt} -  u_yu_{xt})}{u_yu_t}q + f_2q_x
    - \frac{f_2u_x}{u_y}q_y,\\
  w_{21,t}
  &= -\frac{f_2u_t}{u_y}q_y + \frac{f_2u_{yt}}{u_y}q,\\
  w_{22,x}
  &= \nu_{22}q + (f_{2, y}u + 2f_2u_y) q_x
    - \frac{f_2uu_xu_{yy}}{u_y^2}q_y + \frac{f_2uu_x}{u_y}q_{yy},\\
  \intertext{where}
  \nu_{22}
  &= \frac{f_2uu_xu_{ytt}}{u_t^2}
    - \frac{f_2uu_yu_{xtt}}{u_t^2} - \frac{f_2uu_xu_{yyt}}{u_tu_y}
    + \frac{f_2u_y(uu_{tt} - 2u_t^2)u_{xt}}{u_t^3}\\
  &+ \frac{f_2uu_xu_{yt}u_{yy}}{u_tu_y^2}
    - \frac{f_2uu_xu_{yt}u_{tt}}{u_t^3}
    + \frac{2f_2u_xu_{yt}}{u_t} + f_{2, y}u_x,\\
  w_{22,t}
  &= (f_2u_y + f_{2, y})q_t - f_2u_tq_y
    - \frac{f_2uu_tu_{yy}}{u_y^2}q_y
    + \frac{f_2uu_t}{u_y}q_{yy} + f_2uq_{yt}\\
  &+ \left(f_{2, y}u_t + 3f_2u_{yt} - \frac{f_2uu_{yyt}}{u_y}
    + \frac{f_2u_{yy}u_{yt}}{u_y^2}\right)q,\\
  w_{31,x}
  &= -\frac{f_3u_x}{u_t}q_t + f_3q_x + \frac{f_3u_{xt}}{u_t}q,\\
  w_{31,y}
  &= -\frac{f_3u_y}{u_t}q_t + \frac{f_3u_{yt}}{u_t}q.
\end{align*}

\subsection{$A = -C$, $2B = -3C$}\label{sec:a-=-c-9}

The nonlocal variables are defined by
\begin{align*}
  w_{1,y}
  &= \frac{3f_1(2u_xu_{yt} + 3u_yu_xt)}{4\sqrt{u_x}u_t}q
    - \frac{3f_1u_y}{2\sqrt{u_x}}q_x,\\
  w_{1,t}
  &= \frac{3f_1u_{xt}}{2\sqrt{u_x}}q
    - \frac{f_1u_t}{\sqrt{u_x}}q_x + \frac{f_1}{\sqrt{u_x}}q_t,\\
  w_{22,x}
  &= \frac{f_2(2u_xu_{yt} + 3u_yu_{xt})}{2u_yu_t}q
    - \frac{3f_2}{2}q_x - \frac{f_2u_x}{u_y}q_y,\\
  w_{22,t}
  &= -\frac{f_2u_t}{u_y}q_y + \frac{f_2u_{yt}}{u_y}q,\\
  w_{31,x}
  &= -\frac{f_3u_t(2u_xu_{yt} - 3u_yu_{xt})}{3u_y}q
    + 2f_3u_t^2q_x + \frac{2f_3u_t^2u_x}{3u_y}q_y\\
  &+ \frac{f_3u_t(2f_{1, x}u_tu_x + f_1u_tu_{xx} -
    4f_1u_xu_{xt})}{2f_1^2u_x^{3/2}}w_1\\
  &- \frac{f_3u_t^2}{f_1\sqrt{u_x}}w_{1,x}
    - \frac{4f_3u_tu_{xt}}{3f_2}w_{22},\\
  w_{31,y}
  &= \frac{f_3u_t(2u_xu_{yt} - 9u_yu_{xt})}{4u_x}q
    + \frac{3f_3u_t^2u_y}{2u_x}q_x
    + f_3u_t^2q_y\\
  &- \frac{3f_3u_t(2u_xu_{yt} - u_yu_xt)}{4f_1u_x^{3/2}}w_1
    + \frac{2f_3u_t(f_{2, y}u_t - 2f_2u_{yt})}{3f_2^2}w_{22}
    - \frac{2f_3u_t^2}{3f_2}w_{22,y}.
\end{align*}

\subsection{$A = C$, $B = 3C$}\label{sec:a-=-c-10}

The other two recurion operators are
\begin{align*}
  \mathcal{R}_{12}
  &= \nu_{12}q - \frac{f_1u_x^2(8u_tu_{xx} - u_xu_{xt})}{u_t}q_x
    - \frac{f_1u_x^3(5u_yu_tu_{xx} - u_x^2u_{yt})}{3u_tu_y^2}q_y\\
  &+ \frac{f_1u_x^4(u_xu_{yt} + 3u_yu_{xt})}{3u_yu_t^2}q_t
    - 2f_1u_x^3q_{xx} - \frac{f_1u_x^4}{u_t}q_{xt}
    -\frac{f_1u_x^5}{3u_tu_y}q_{yt}\\
  &- \frac{2f_{1, x}u_x}{f_1}w_{1,x}
    + \frac{4f_{1, x}u_x}{9f_1}w_{11,x} + u_xw_{1,xx} - \mu_{12}w_{22}
    - \frac{2u_x}{9}w_{11,xx},
    \intertext{where}
  \nu_{12}
  &= -2f_1u_x^2u_{xxx} + \frac{f_1u_x^3u_{xxt}}{u_t}
    + \frac{f_1u_x^4u_{xtt}}{u_t^2} + \frac{f_1u_x^5u_{ytt}}{3u_yu_t^2}\\
  &- 2f_1u_xu_{xx}^2
    + \frac{f_1u_x^2(5u_xu_{yt} + 12u_yu_{xt})u_{xx}}{3u_tu_y}
    - \frac{f_1u_x^4(u_xu_yt + 3u_yu_{xt})u_{tt}}{3u_yu_t^3}\\
  &- \frac{f_1u_x^3u_{xt}^2}{u_t^2}
    - \frac{f_1u_x^5u_{yt}^2}{3u_t^2u_y^2}
    + \frac{u_x^3(2f_{1, x}^2 - f_1f_{1, x, x})}{f_1},\\
  \mu_{12}
  &= -\frac{2f_1u_x^2u_{xxx}}{3f_2} - \frac{2f_1u_xu_{xx}^2}{3f_2}
    + \frac{(2f_{1, x}^2 - f_1f_{1, x, x})u_x^3}{3f_1f_2},\\
  \mathcal{R}_{32}
  &= \left(\frac{u_t^3u_{ytt}}{2u_y} - \frac{u_t^3u_{yt}^2}{2u_y^2}
    + \frac{2u_t^2u_{tt}u_{yt}}{u_y}\right) q
    +\frac{u_t^3(u_tu_{yt} - 5u_yu_{tt})}{2u_y^2} q_y
    + \frac{u_t^3u_{yt}}{2u_y}q_t\\
  &- \frac{u_t^4}{2u_y}q_{yt}
    - \left(\frac{u_t^3(2f_{3, t}^2 - f_{3, t, t}f_3)
    }{2f_2f_3^2}
 - \frac{u_t^2u_{ttt}}{f_2} - \frac{u_tu_{tt}^2}{f_2}
    \right)w_{22}\\
  &- \frac{2u_tf_{3, t}}{3f_3^2}w_{31,t} + \frac{u_t}{3f_3}w_{31,tt}.
\end{align*}
with the defining equations
\begin{align*}
  w_{1,y}
  &= \frac{3f_1u_x(u_xu_{yt} + 3u_yu_{xt})}{u_t}q - 3f_1u_yu_xq_x,\\
  w_{1,t}
  &= 3f_1u_xu_{xt}q + f_1q_tu_x^2 - f_1u_tu_xq_x,\\
  w_{11,y}
  &=\frac{9f_1u_x(u_xu_{yt} + 3u_yu_{xt})}{2u_t}q - \frac{27f_1u_yu_x}{2q_x}
    - \frac{9f_1u_x^2}{2}q_y\\
  &- \frac{3f_1u_x(f{2, y}u_xu_t - 2f_2u_xu_{yt} -
    6f_2u_yu_{xt})}{2f_2^2u_t}w_{22} + \frac{3f_1u_x^2}{2f_2}w_{22,y},\\ 
  w_{11,t}
  &= \frac{3f_1u_x(u_xu_{yt} + 3u_yu_{xt})}{2u_y}q
    - \frac{9f_1u_tu_x}{2}q_x
    - \frac{3f_1u_x^2u_t}{2u_y}q_y + \frac{3f_1u_xu_{xt}}{f_2}w_{22},\\
  w_{22,x}
  &= \frac{f_2(u_xu_{yt} + 3u_yu_{xt})}{u_yu_t}q - 3 f_2q_x
    - \frac{f_2u_x}{u_y}q_y,\\
  w_{22,t}
  &= -\frac{f_2u_t}{u_y}q_y + \frac{f_2u_{yt}}{u_y}q,\\
  w_{31,x}
  &= -\frac{3f_3u_t(u_xu_{yt} + 3u_yu_{xt}}{2u_y}q
    + \frac{9f_3u_t^2}{2}q_x + \frac{3f_3u_t^2u_x}{2u_y}q_y
    - \frac{3f_3u_tu_{xt}}{f_2}w_{22},\\
  w_{31,y}
  &= \frac{3f_3u_t(f_{2, y}u_t - 2f_2u_{yt})}{2f_2^2}w_{22}
    - \frac{3f_3u_t^2}{2f_2}w_{22,y}.
\end{align*}

\subsection{$A = -C$, $2B = 3C$}\label{sec:a-=-c-11}

These are the recusion operators
\begin{align*}
  \mathcal{R}_1
  &= \frac{u_x^2(4u_x^2{u_yt} - 24u_tu_yu_{xx} +
    3u_xu_yu_xt)}{12u_yu_t} q - \frac{u_x^4}{3u_y}q_y -
    \frac{u_x^4}{2u_t}q_t,\\
  &-
  \frac{4f_{1, x}u_x}{9f_1^2}w_{11} + \frac{4u_x}{9f_1}w_{11,x} +
    \frac{2u_x^2u_{xx}}{3f_2}w_{22}
    + \frac{(4u_tu_{xx} - u_xu_{xt})u_x^2}{4f_3u_t^{3/2}}w_3,\\
  \mathcal{R}_2
  &= \nu_2q + \frac{u_x^3(5u_tu_yu_{xx} - u_x^2u_{yt})}{u_tu_y^2}q_y +
  \left(\frac{15u_x^3u_{xx}}{2u_t} - \frac{3u_x^4u_{xt}}{4u_t^2} -
    \frac{u_x^5u_{yt}}{u_t^2u_y}\right)q_t\\
  &+ \frac{u_x^5}{u_yu_t}q_{yt}
    - \frac{4(2f_{1, x}^2 - f_{1, x, x}f_1)u_x}{3f_1^3}w_{11}
    + \frac{8f_{1, x}u_x}{3f_1^2}w_{11,x}\\
  &- \frac{4u_x}{3f_1}w_{11,xx}
    - \frac{2(u_xu_{xxx} + u_{xx}^2)u_x}{f_2} w_{22}-\mu_2 w3,
         \intertext{where}
    \nu_2
  &= \frac{6u_x^2u_{xxx} - 3u_x^3u_{xxt}}{4u_t}
    + \frac{3u_x^4u_{xtt}}{2u_t^2} -
    \frac{u_x^5u_{ytt}}{u_t^2u_y} + 6u_xu_{xx}^2 +
    \frac{15u_x^3u_{xt}^2}{8u_t^2}\\ 
  &- \frac{u_x^2(5u_xu_{yt} + 3u_yu_{xt})u_{xx}}{u_tu_y}
     -
  \frac{3u_x^4u_{tt}u_{xt}}{2u_t^3} + \frac{u_x^5u_{yt}u_{tt}}{u_t^3u_y} +
    \frac{u_x^5u_{yt}^2}{u_t^2u_y^2}
    \intertext{and}
  \mu_2
  &= \frac{3u_x^2u_{xxx}}{f_3\sqrt{u_t}} - \frac{3u_x^3u_{xxt}}{4f_3u_t^{3/2}}
    + \frac{3u_x(8u_t^2u_{xx}^2 - 8u_tu_xu_{xt}u_{xx} +
    3u_x^2u_{xt}^2)}{8f_3u_t^{5/2}}. 
\end{align*}
and here are the defining equations
\begin{align*}
  w_{11,y}
  &= \frac{9f_1u_x(5u_xu_{yt} - 12u_yu_xt)}{16u_t}q + \frac{9f_1u_x^2}{4}q_y
    + \frac{27f_1u_x^2u_y}{16u_t}q_t\\
  &+ \frac{3f_1u_x(f_{2, y}u_tu_x - 2f_2u_xu_{yt} +
    3f_2u_yu_xt)}{4f_2^2u_t}w_{22} 
    - \frac{3f_1u_x^2}{4f_2}w_{22,y}\\
  &- \frac{27f_1u_x(u_xu_{yt} - 2u_yu_{xt})}{16f_3u_t^{3/2}}w_3,\\
  w_{11,t}
  &= -\frac{3f_1u_x(u_xu_{yt} - 6u_yu_{xt})}{4u_y}q
    + \frac{3f_1u_x^2u_t}{4u_y}q_y + \frac{9f_1u_x^2}{4}q_t
    - \frac{3f_1u_xu_{xt}}{2f_2}w_{22}\\
  &+ \frac{9f_1u_x(2f_{3, t}u_tu_x - 4f_3u_tu_{xt} +
    f_3u_xu_{tt})}{16f_3^2u_t^{3/2}}w_3 
    - \frac{9f_1u_x^2}{8f_3\sqrt{u_t}}w_{3,t},\\
  w_{22,x}
  &= \frac{f_2(2u_xu_{yt} - 3u_yu_{xt})}{2u_yu_t} q
    + \frac{3f_2}{2}q_x - \frac{f_2u_x}{u_y}q_y,\\
  w_{22,t}
  &= -\frac{f_2u_t}{u_y}q_y + \frac{f_2u_{yt}}{u_y}q,\\
  w_{3,x}
  &= \frac{3f_3u_{xt}}{2\sqrt{u_t}}q + f_3\sqrt{u_t}q_x
    - \frac{f_3u_x}{\sqrt{u_t}}q_t,\\
  w_{3,y}
  &= \frac{3f_3u_{yt}}{2\sqrt{u_t}}q - \frac{3f_3u_y}{2\sqrt{u_t}}q_t.
\end{align*}

\subsection{$A = -C$, $B = 3C$}\label{sec:a-=-c-12}

These are the rest four operators:
\begin{align*}
  \mathcal{R}_{12}
  &= 3u_x^2\left(\frac{f_1u_x^2u_{yt}}{2u_tu_y} - 3f_1u_{xx}
    - \frac{3f_1u_xu_{xt}}{2u_t}  - \frac{3f_{1, x}u_x}{2}\right)q -
    \frac{3f_1u_x^4}{2u_y}q_y\\
  &+ u_xw_{11,x}
    + \frac{3u_x^2(f_{1, x}u_x + 2f_1u_{xx})}{2f_2}w_{22},\\
  \mathcal{R}_{13}
  &= \nu_{13}q + \frac{u_x^3u_{xt}}{2u_t}q_x +
    u_x^3\left(\frac{5u_{xx}}{6u_y} - \frac{u_x^2u_{yt}}{6u_tu_y^2}
    + \frac{f_{1, x}u_x}{3f_1u_y}\right)q_y\\
  &- \frac{u_x^4(u_xu_{yt} - 3u_yu_{xt})}{6u_t^2u_y}q_t
    - \frac{u_x^4}{2u_t}q_{xt} + \frac{u_x^5}{6u_tu_y}q_{yt}\\
  &- \left(\frac{u_x^2u_{xxx}}{3f_2} + \frac{u_xu_{xx}^2}{3f_2} +
  \frac{2f_{1,x}u_x^2u_{xx}}{3f_1f_2} +
    \frac{u_x^3f_{1,x,x}}{6f_1f_2}\right)w_{22} - \frac{u_x}{9f_1}w_{11,xx},
    \intertext{where}
  \nu_{13}
  &= u_x^2u_{xxx} + \frac{u_x^3u_{xxt}}{2u_t} + \frac{u_x^4u_{xtt}}{2u_t^2} -
    \frac{u_x^5u_{ytt}}{6u_t^2u_y}
    + \frac{u_x^4(u_xu_{yt} - 3u_yu_{xt})u_{tt}}{6u_yu_t^3}\\
  &+ \frac{u_x^5u_{yt}^2}{6u_t^2u_y^2}
    - \frac{u_x^3(2f_{1, x}u_x + 5f_1u_{xx}) u_yt}{6f_1u_tu_y}
    - \frac{u_x^3u_{xt}^2}{2u_t^2}
    + \frac{u_x^2(f_{1, x}u_x + 2f_1u_{xx})u_{xt}}{f_1u_t}\\
  &+ u_xu_{xx}^2 + \frac{2f_{1, x}u_x^2u_{xx}}{f_1}
  + \frac{u_x^3f_{1, x, x}}{2f_1},\\
  \mathcal{R}_{32}
  &= -\frac{u_t^2(6f_{3, t}u_tu_y + 12f_3u_yu_{tt} - f_3u_tu_{yt})}{3u_y}q
    - \frac{f_3u_t^4}{3u_y}q_y - 2f_3u_t^3q_t\\
  &+ \frac{u_t^2(f_{3, t}u_t + 2f_3u_{tt})}{3f_2}w_{22} + u_tw_{3,t} -
  \frac{2u_t}{9}w_{31,t},\\
  \mathcal{R}_{33}
  &= \nu_{33}q -
    \frac{u_t^3(2f_{3, t}u_tu_y - f_3u_tu_{yt} + 5f_3u_yu_{tt})}{3u_y^2}q_y\\
  &- u_t^2\left(8f_3u_{tt} - \frac{f_3u_tu_{yt}}{3u_y}
    + 4f_{3, t}u_t\right)q_t - 2f_3u_t^3q_{tt}
    - \frac{f_3u_t^4}{3u_y}q_{yt}\\
  &+
  \frac{u_t(4f_{3, t}u_tu_{tt}  + 2f_3u_tu_{ttt} + f_{3, t, t}u_t^2 +
    2f_3u_{tt}^2)}{3f_2}w_{22} - \frac{2u_t}{9}w_{31,tt} + u_tw_{3,tt},
    \intertext{where}
  \nu_{33}
  &= \frac{f_3u_t^3u_{ytt}}{3u_y} - 4f_3u_t^2u_{ttt}
    - \frac{f_3u_t^3u_{yt}^2}{3u_y^2}
    + \frac{4f_3u_t^2u_{tt}}{3u_y}
    + \frac{2f_{3, t}u_t^3u_{yt})}{3u_y}\\
  &- 4f_3u_tu_{tt}^2 - 8f_{3, t}u_t^2u_{tt} - 2f_{3, t, t}u_t^3.
\end{align*}
The defining equations are
\begin{align*}
  w_{11,y}
  &= \frac{9f_1u_x(u_xu_{yt} - 3u_yu_{xt})}{u_t}q
    + \frac{9f_1u_x^2}{2}q_y\\
  &+ \frac{3f_1u_x(f_{2, y}u_xu_t - 2f_2u_xu_{yt} +
    6f_2u_yu_{xt})}{2f_2^2u_t}w_{22}  - \frac{3f_1u_x^2}{2f_2}w_{22,y},\\
  w_{11,t}
  &= -\frac{3f_1u_x(u_xu_{yt} - 6u_yu_{xt})}{2u_y}q +
    \frac{9f_1u_x^2}{2}q_t
    + \frac{3f_1u_tu_x^2}{2u_y}q_y - \frac{3f_1u_xu_{xt}}{f_2}w_{22},\\
  w_{22,x}
  &= \frac{f_2(u_xu_{yt} - 3u_yu_{xt})}{u_yu_t}q + 3f_2q_x -
    \frac{f_2u_x}{u_y}q_y,\\
  w_{22,t}
  &= -\frac{f_2u_t}{u_y}q_y + \frac{f_2u_{yt}}{u_y}q,\\
  w_{3,x}
  &= 3f_3u_tu_{xt}q - f_3u_tu_xq_t + f_3u_t^2q_x,\\
  w_{3,y}
  &= 3f_3u_tu_{yt}q - 3f_3u_tu_yq_t,\\
  w_{31,x}
  &= \frac{3f_3u_t(u_xu_{yt} - 6u_yu_{xt})}{2u_y}q
    - \frac{9f_3u_tu_x}{2}q_t
    - \frac{3f_3u_t^2u_x}{2u_y}q_y + \frac{3f_3u_tu_{xt}}{f_2}w_{22},\\
  w_{31,y}
  &= -\frac{9f_3u_tu_{yt}}{2}q
    - \frac{27f_3u_tu_y}{2}q_t - 9f_3u_t^2q_y\\
  &- \frac{3f_3u_t(f_{2, y}u_t - 2f_2u_{yt})}{2f_2^2}w_{22}
    + \frac{3f_3u_t^2}{2f_2}w_{22,y}.
\end{align*}

\subsection{$3A = 2C$, $3B = -2C$}\label{sec:3a-=-2c-1}

The second and the third operators look as follows:
\begin{align*}
  \mathcal{R}_2
  &= -\frac{2f_2u_y^4u_{xt}}{3u_xu_t}q
    + \frac{f_2u_y^4}{u_x}q_x
    + \frac{2f_2u_y^4}{3u_t}q_t\\
  &- \left(\frac{3f_2u_{yy}}{f_1\sqrt{u_x}} -
    \frac{f_2u_y^2u_{xt}}{2f_1u_x^{3/2}u_t} 
    + \frac{f_2u_yu_{yt}}{2f_1\sqrt{u_x}u_t}
    + \frac{3f_{2, y}u_y}{2f_1\sqrt{u_x}}\right)u_y^2w_1 + u_yw_{21,y}\\
  &+ \frac{u_y^2(2f_2u_{yy} + f_{2, y}u_y)}{f_3}w_{31},\\
  \mathcal{R}_3
  &= \nu_3q + \left(-\frac{f_2u_y^5u_{xt}}{u_x^2u_t}
    + \frac{5f_2u_y^3u_{yy}}{u_x} +\frac{f_2u_y^4u_{yt}}{3u_xu_t}
    + \frac{2f_{2, y}u_y^4}{u_x}\right)q_x\\
  &- \left(\frac{10f_2u_y^3u_{yy}}{3u_t} - \frac{2f_2u_y^5u_{xt}}{3u_xu_t^2}
    + \frac{4f_{2, y}u_y^4}{3u_t}\right)q_t 
    + \frac{2f_2u_y^5}{3u_xu_t}q_{xt} + \mu_3w_1\\
  &+\frac{u_y(2f_2u_yu_{yyy} +
    f_{2, y, y}u_y^2 + 4f_{2, y}u_yu_{yy} + 2f_2u_{yy}^2)}{f_3}w_{31} +
    u_yw_{21,yy}, 
    \intertext{where}
    \nu_3
  &= -\frac{4f_2u_y^5u_{xtt}}{9u_xu_t^2}
    + \frac{4f_2u_y^4u_{ytt}}{9u_t^2}
    - \frac{4f_2u_y^4(u_xu_{yt} - u_yu_{xt})u_{tt}}{9u_xu_t^3}
    -\frac{10f_2u_y^3u_{xt}u_{yy}}{3u_tu_x}\\
  &+ \frac{2f_2u_y^5u_{xt}^2}{3u_t^2u_x^2}
    - \frac{2u_y^4(6f_{2, y}u_t + f_2u_{yt})u_{xt}}{9u_xu_t^2}
    + \frac{2f_2u_y^3u_yt^2}{9u_t^2}
    \intertext{and}
    \mu_3
  &= \frac{f_2u_y^5u_{xtt}}{3f_1u_t^2u_x^{3/2}}
    - \frac{3f_2u_y^2u_{yyy}}{f_1\sqrt{u_x}}
    - \frac{f_2u_y^3u_{yyt}}{2f_1u_t\sqrt{u_x}}
    - \frac{f_2u_y^4u_{ytt}}{3f_1u_t^2\sqrt{u_x}}\\
  &- \frac{3f_2u_yu_{yy}^2}{f_1\sqrt{u_x}}
    + \left(\frac{5f_2u_y^3u_{xt}}{2f_1u_tu_x^{3/2}} -
    \frac{2f_2u_y^2u_{yt}}{f_1u_t\sqrt{u_x}} 
    - \frac{6f_{2, y}u_y^2}{f_1\sqrt{u_x}}\right)u_{yy}\\
  &+ \frac{f_2u_y^3u_{yt}^2}{3f_1u_t^2\sqrt{u_x}}
    + \frac{f_2u_y^4u_{xt}}{6f_1u_t^2u_x^{3/2}}
    + \frac{f_2u_y^4u_{tt}}{3f_1\sqrt{u_x}u_t^3}
    - \frac{f_{2, y}u_y^3u_{yt}}{f_1u_t\sqrt{u_x}}
    - \frac{f_2u_y^5u_{xt}^2}{2f_1u_t^2u_x^{5/2}}\\
  &+ \frac{u_y^4(3f_{2, y}u_t^2 - f_2y_yu_{tt})u_{xt}}{3f_1u_x^{3/2}u_t^3}
    - \frac{3f_{2, y, y}u_y^3}{2f_1\sqrt{u_x}}.
\end{align*}
Nonlocalities are defined by the equations
\begin{align*}
  w_{1,y}
  &= -\frac{4f_1(u_xu_{yt} - u_yu_{xt})}{9u_t\sqrt{u_x}}q
    - \frac{2f_1u_y}{3\sqrt{u_x}}q_x,\\
  w_{1,t}
  &= \frac{2f_1u_{xt}}{3\sqrt{u_x}}q - \frac{f_1u_t}{\sqrt{u_x}}q_x
    - \frac{2f_1\sqrt{u_x}}{3}q_t,\\
  w_{21,x}
  &= \frac{2f_2u_y^2u_{xt}}{3u_t}q - f_2u_y^2q_x
    - \frac{2f_2u_y^2u_x}{3u_t}q_t\\
  &- \frac{f_2u_y(8f_1u_tu_x^{7/2}u_{yt} + (6f_{1, x}u_tu_y -
    8f_1u_yu_{xt})u_tu_x^{5/2} + 3f_1u_t^2u_yu_x^{3/2}u_{xx})}{4f_1^2}w_1\\
  &+ \frac{3f_2u_y^2}{2f_1\sqrt{u_x}}w_{1,x}
    + \frac{4f_2u_y(u_xu_{yt} - u_yu_{xt})}{3f_3u_t}w_{31},\\
  w_{21,t}
  &= \frac{f_2u_y^2u_{xt}}{u_x}q - \frac{3f_2u_y^2u_t}{2u_x}q_x
    - f_2u_y^2q_t + \frac{3f_2u_y(4u_xu_{yt} - u_yu_{xt})}{4f_1u_x^{3/2}}w_1\\
  &+ \frac{f_2u_y(f_{3, t}u_y - 2f_3u_{yt})}{f_3^2}w_{31}
    - \frac{f_2u_y^2}{f_3}w_{31,t},\\
  w_{31,x}
  &= \frac{2f_3u_x}{3u_t}q_t + f_3q_x - \frac{2f_3u_{xt}}{3u_t}q,\\
  w_{31,y}
  &= \frac{2f_3u_y}{3u_t}q_t - \frac{2f_3u_{yt}}{3u_t}q.
\end{align*}

\subsection{$3A = C$, $3B = -C$}\label{sec:3a-=-c-1}

One has
\begin{align*}
  \mathcal{R}_{12}
  &= \nu_{12}w_{31} + \left(\frac{2f_1uu_x^2u_{xx}}{f_3u_t}
    - \frac{f_1uu_x^3u_{xt}}{f_3u_t^2}
    + \frac{f_1u_x^4}{f_3u_t}
    + \frac{f_{1, x}uu_x^3}{f_3u_t}\right)w_{31,t}\\
  &+ \left(\frac{4f_1u_x^2u_{xx}}{f_3u_t}
    - \frac{2f_1u_x^3u_{xt}}{f_3u_t^2}
    + \frac{2f_{1, x}u_x^3}{f_3u_t}\right)w_{32} + 3u_xw_{11,x},
    \intertext{where}
  \nu_{12}
  &= \frac{f_{3, t}f_1uu_x^3u_{xt}}{f_3^2u_t^2}
    - \frac{2f_1u_x^2(f_3u_t + f_{3, t}u)u_{xx}}{f_3^2u_t}
    - \frac{f_{3, t}f_1u_x^4}{f_3^2u_t}\\
  &- \frac{f_{1, x}(f_3u_t + f_{3, t}u)u_x^3}{f_3^2u_t},\\
  \mathcal{R}_{13}
  &= \frac{f_1u_x^3(u_tu_xu_{xtt} - u_tu_xt^2 - u_xu_{tt}u_{xt})}{9u_t^3}q
    - \frac{f_1u_x^3u_{xt}}{3u_t}q_x\\
  &+ \frac{f_1u_x^4(u_tu_{xt} + u_x)u_{tt}}{9u_t^3}q_t
    - \frac{f_1u_x^5}{9u_t^2}q_{tt} + \frac{f_1u_x^4}{3u_t}q_{xt}
    - 2\frac{f_{1, x}u_x}{f_1}w_{11,x}\\
  &+ \nu_{13}w_{31} + \mu_{13}w_{31,t} - \lambda_{13}w_{32} + u_xw_{11,xx},
    \intertext{where}
  \nu_{13}
  &= -\frac{2f_1u_x^2(f_3u_t + f_{3, t}u)u_{xxx}}{3f_3^2u_t}
    + \frac{f_{3, t}f_1uu_x^3u_{xxt}}{f_3^2u_t^2}
    - \frac{2f_{3, t}f_1uu_x^3u_{xt}^2}{3f_3^2u_t^3}\\
  &+ \frac{2f_{3, t}f_1u_x^2(2uu_{xx} + u_x^2)u_{xt}}{3f_3^2u_t^2}
    - \frac{2f_1u_x(f_3u_t + f_{3, t}u)u_{xx}^2}{3f_3^2u_t}\\
  &- \frac{5f_{3, t}f_1u_x^3u_{xx}}{3f_3^2u_t}
    + \frac{u_x^3(2f_{1, x}^2 - f_{1, x, x}f_1)(f_3u_t + f_{3,
    t}u)}{3f_3^2f_1u_t},\\ 
  \mu_{13}
  &= \frac{2f_1uu_x^2u_{xxx}}{3f_3u_t}
    -\frac{f_1uu_x^3u_{xxt}}{3f_3u_t^2}
    + \frac{2f_1uu_x^3u_{xt}^2}{3f_3u_t^3}
    - \frac{2f_1u_x^2(2uu_{xx} + u_x^2)u_{xt}}{3f_3u_t^2}\\
  &+ \frac{2f_1uu_xu_{xx}^2}{3f_3u_t}
    + \frac{5f_1u_x^3u_{xx}}{3f_3u_t}
    - \frac{uu_x^3(2f_{1, x}^2 - f_{1, x, x}f_1)}{3f_3f_1u_t},\\
  \lambda_{13}
  &= -\frac{4f_1u_x^2u_{xxx}}{3f_3u_t}
    + \frac{2f_1u_x^3u_{xxt}}{3f_3u_t^2}
    - \frac{4f_1u_x^3u_{xt}^2}{3f_3u_t^3}
    + \frac{8f_1u_x^2u_{xx}u_{xt}}{3f_3u_t^2}\\
  &- \frac{4f_1u_xu_{xx}^2}{3f_3u_t}
    + \frac{2u_x^3(2f_{1, x}^2 - f_{1, x, x}f_1)}{3f_3f_1u_t},\\
  \mathcal{R}_{22}
  &= \frac{u_y^2(6f_{2, y}u_yu_t - f_2u_yu_{yt} + 12f_2u_tu_{yy})}{12f_2u_t}q
    + \frac{u_y^3}{2}q_y + \frac{u_y^4}{12u_t}q_t -
    \frac{3u_y}{4f_2}w_{2,y}\\
  &+ \frac{u_y}{4f_2}w_{21,y}
    - \frac{u_y^2(f_{2, y}u_y + 2f_2u_{yy})}{4f_2f_3}w_{31},\\
  \mathcal{R}_{23}
  &= \nu_{23}q + \frac{f_2u_y^2(24u_tu_{yy} - u_yu_{yt})}{3u_t}q_y
    + \frac{f_2u_y^3(5u_tu_{yy} - u_yu_{yt})}{3u_t^2}q_t + 2f_2u_y^3q_{yy}\\
  &+ \frac{f_2u_y^4}{3u_t}q_{yt} + \frac{6f_{2, y}u_y}{f_2}w_{2,y}
    - 3u_yw_{2,yy} - \frac{2f_{2, y}u_y}{f_2}w_{21,y} + u_yw_{21,yy}\\
  &+ \left(-\frac{2f_2u_y^2u_{yyy}}{f_3}
    - \frac{2f_2u_yu_{yy}^2}{f_3}
    + \frac{(2f_{2, y}^2 - f_{2, y, y}f_2) u_y^3}{f_2f_3}\right)w_{31},
    \intertext{where}
  \nu_{23}
  &= 4f_2u_y^2u_{yyy} - \frac{f_2u_y^3u_{yyt}}{3u_t}
    + \frac{f_2u_y^3u_{yt}^2}{3u_t^2}
    - \frac{4f_2u_y^2u_{yy}u_{yt}}{3u_t}\\
  &+ 4f_2u_yu_{yy}^2 - \frac{2u_y^3(2f_{2, y}^2 - f_{2, y, y}f_2)}{f_2}.
\end{align*}
The defining equations are
\begin{align*}
  w_{11,y}
  &= 
    - \left(\frac{f_1u_x^2(f_{3, t}u - 2f_3u_t)u_{yt}}{9f_3^2u_t^2}
    + \frac{2f_1u_xu_y(f_3u_t + f_{3, t}u)u_{xt}}{9f_3^2u_t^2}
    - \frac{f_{3,  t}f_1u_yu_x^2}{3f_3^2u_t}\right)w_{31}\\
  &-\frac{f_1u_x^2u_{yt}}{3u_t}q
    + \frac{f_1u_x(uu_xu_{yt} + 2uu_yu_{xt} - 3u_tu_xu_y)}{9f_3u_t^2}w_{31,t}\\
  &+ \frac{2f_1u_x(u_xu_{yt} + 2u_yu_{xt})}{9f_3u_t^2}w_{32}
    + \frac{f_1u_x^2u_y}{3u_t}q_t,\\
  w_{11,t}
  &= \nu_{11}w_{31} +
    \frac{f_1uu_x(2f_{3, t}u_xu_t - 2f_3u_tu_{xt} + f_3u_xu_{tt})}{3f_3^2u_t^2}
    w_{31,t}
    - \frac{f_1uu_x^2}{3f_3u_t}w_{31,tt}\\
  &+ \frac{2f_1u_x(f_{3, t}u_xu_t - 2f_3u_tu_{xt} +
    f_3u_xu_{tt})}{3f_3^2u_t^2}w_{32} - \frac{2f_1 u_x^2}{3f_3u_t}w_{32,t}, 
    \intertext{where}
    \nu_{11}
  &= -\frac{f_1f_{3, t}uu_x^2u_{tt}}{3f_3^2u_t^2} +
    \frac{2f_1u_x(f_3u_t + f_{3, t}u)u_{xt}}{3f_3^2u_t} -
    \frac{f_1u(2f_{3, t}^2 - f_{3, t, t}f_3) u_x^2}{3f_3^3u_t},\\
  w_{2,x}
  &= \frac{f_2u_y(u_xu_{yt} - u_yu_{xt})}{3u_t}q
    + \frac{f_2u_y^2}{3}q_x - \frac{f_2u_xu_y}{3}q_y,\\
  w_{2,t}
  &= f_2u_yu_{yt}q - f_2u_tu_yq_y,\\
  w_{21,x}
  &= -\frac{f_2u_y(u_xu_{yt} - 2u_yu_{xt})}{3u_t}q +
    \frac{2f_2u_y(u_xu_{yt} - u_yu_{xt})}{3f_3u_t}w_{31}\\
  &- \frac{f_2u_y^2u_x}{3u_t}q_t - f_2u_xu_yq_y,\\
  w_{21,t}
  &= -f_2u_yu_{yt}q - \frac{f_2u_y(f_{3, t}u_y - 2f_3u_{yt})}{f_3^2}w_{31} -
    2f_2u_y^2q_t\\
  &- 3f_2u_tu_yq_y + \frac{f_2u_y^2}{f_3}w_{31,t},\\
  w_{31,x}
  &= -\frac{f_3u_x}{3u_t}q_t + f_3q_x + \frac{f_3u_{xt}}{3u_t}q,\\
  w_{31,y}
  &= -\frac{f_3u_y}{3u_t}q_t + \frac{f_3u_{yt}}{3u_t}q,\\
  w_{32,x}
  &= -\frac{f_3(uu_tu_{xtt} - uu_{tt}u_{xt} - u_t^2u_{xt})}{6u_t^2}q -
    \frac{f_3u_x(uu_{tt} + u_t^2)}{6u_t^2}q_t\\
  &+ \frac{f_3u_t}{2}q_x + \frac{f_3uu_x}{6u_t}q_{tt} - \frac{f_3u}{2}q_{xt},\\
  w_{32,y}
  &= -\frac{f_3(uu_tu_{ytt} - uu_{tt}u_{yt} - 4u_t^2u_{yt})}{6u_t^2}q -
    \frac{f_3u_y(uu_{tt} + 4u_t^2)}{6u_t^2}q_t + \frac{f_3uu_y}{6u_t}q_{tt}.
\end{align*}

\subsection{$A = C$, $2B = -C$}\label{sec:a-=-c-13}

We have
\begin{align*}
  \mathcal{R}_3
  &= \nu_2q
    - \frac{u_y^3(20u_tu_xu_{yy} + 2u_xu_yu_{yt} -
    3u_y^2u_{xt})}{16u_tu_x^2}q_x 
    - \frac{u_y^3u_{yt}}{4u_t}q_y\\
  &- \frac{u_y^4(2u_xu_{yt} - u_yu_{xt})}{8u_t^2u_x}q_t
    - \frac{u_y^5}{8u_tu_x}q_{xt} + \frac{u_y^4}{4u_t}q_{yt} - \mu_2w_1\\
  &+ \frac{u_y(2f_{2, y}^2 - f_{2, y, y}f_2)}{f_2^3}w_{21}
    - \frac{2f_{2, y}u_y}{f_2^2}w_{21,y} + \frac{u_y}{f_2}w_{21,yy},
    \intertext{where}
  \nu_2
  &= \frac{u_y^5u_{xtt}}{16u_t^2u_x} - \frac{u_y^3u_{yyt}}{4u_t}
    - \frac{u_y^4u_{ytt}}{8u_t^2}
    + \frac{u_y^4(2u_xu_{yt} - u_yu_{xt})u_{tt}}{16u_xu_t^3}\\
  &- \frac{u_y^2(8u_xu_{yt} - 5u_yu_{xt})u_{yy}}{8u_tu_x}
    + \frac{u_y^3u_{yt}^2}{8u_t^2} + \frac{u_y^4u_{xt}u_{yt}}{16u_t^2u_x}
    - \frac{3u_y^5u_{xt}^2}{32u_t^2u_x^2},\\
  \mu_2
  &= \frac{u_y^5u_{xtt}}{16f_1u_t^2u_x^{3/2}} -
    \frac{u_y^2u_{yyy}}{f_1\sqrt{u_x}}
    - \frac{u_y^3u_{yyt}}{4f_1u_t\sqrt{u_x}}
    - \frac{u_y^4u_{ytt}}{8f_1u_t^2\sqrt{u_x}}\\
  &+ \frac{u_y^4(2u_xu_{yt} - u_yu_{xt})u_{tt}}{16f_1u_x^{3/2}u_t^3}
    - \frac{3u_y^5u_{xt}^2}{32f_1u_t^2u_x^{5/2}}
    + \frac{u_y^3(10u_tu_{yy} + u_yu_{yt})u_{xt}}{16f_1u_t^2u_x^{3/2}}\\
  &- \frac{u_yu_{yy}^2}{f_1\sqrt{u_x}}
    - \frac{u_y^2u_{yt}u_{yy}}{f_1u_t\sqrt{u_x}}
    + \frac{u_y^3u_{yt}^2}{8f_1u_t^2\sqrt{u_x}},
\end{align*}
and
\begin{align*}
  w_{1,y}
  &= -\frac{f_1(2u_xu_{yt} - u_yu_{xt})}{4u_t\sqrt{u_x}}q
    - \frac{f_1u_y}{2\sqrt{u_x}}q_x,\\
  w_{1,t}
  &= \frac{f_1u_{xt}}{2\sqrt{u_x}}q - \frac{f_1u_t}{\sqrt{u_x}}q_x
    - f_1\sqrt{u_x}q_t ,\\
  w_{21,x}
  &= \left(\frac{f_2u_y^2u_{xx}}{4f_1u_x^{3/2}}
    - \frac{f_2u_y^2u_{xt}}{2f_1sqrt{u_x}u_t}
    + \frac{f_2u_y\sqrt{u_x}u_{yt}}{f_1u_t}
    + \frac{f_2u_y^2f_{1, x}}{2f_1^2\sqrt{u_x}}\right)w_1\\
  &- \frac{f_2u_y^2}{2f_1\sqrt{u_x}}w_{1,x},\\
  w_{21,t}
  &= -\frac{f_2u_y^2u_{xt}}{4u_x}q + \frac{f_2u_y^2u_t}{2u_x}q_x
    + \frac{f_2u_y^2}{2}q_t
    - \frac{f_2u_y(4u_xu_{yt} - u_yu_{xt})}{4f_1u_x^{3/2}}w_1.
\end{align*}
The nonlocal variables that enter the expression fo~$\mathcal{I}$ are defined
by
\begin{align*}
  w_{22,x}
  &= -\frac{f_1\sqrt{u_x}(2u_xu_{yt} - u_yu_{xt})}{2u_yu_t^2}p
    - \frac{f_1\sqrt{u_x}}{u_t}p_x
    + \frac{2f_1u_x^{3/2}}{u_tu_y}p_y\\
  &+ \left(\frac{u_{xx}}{2u_x} - \frac{u_{xt}}{u_t}
    + \frac{2u_xu_{yt}}{u_tu_y} + \frac{f_{1, x}}{f_1}\right)w_{22},\\
  w_{22,t}
  &= \frac{f_1\sqrt{u_x}u_{yt}}{u_yu_t}p
    - \frac{2f_1\sqrt{u_x}}{u_y}p_y
    - \frac{4u_xu_{yt} - u_yu_{xt}}{2u_xu_y}w_{22},\\
  w_{23,x}
  &= \frac{f_3u_{xt}}{u_t^{3/2}}p
    - \frac{2f_3u_x}{\sqrt{u_t}u_y}p_y
    - \frac{3f_3(2u_xu_{yt} - u_yu_xt)}{2f_1\sqrt{u_t}u_y\sqrt{u_x}}w_{22}
    - \frac{2f_3\sqrt{u_t}\sqrt{u_x}}{f_1u_y}w_{22,y}\\
  &- \frac{u_xu_{yt} - u_yu_{xt}}{u_tu_y}w_{23},\\
  w_{23,t}
  &= -\frac{f_3(u_tu_{yt} - u_yu_{tt})}{u_t^{3/2}u_y}p
    - \frac{f_3}{\sqrt{u_t}}p_t
    +\frac{f_3\sqrt{u_t}(10u_xu_{yt} - u_yu_{xt})}{2f_1u_yu_x^{3/2}}w_{22}\\
  &+ \frac{2f_3u_t^{3/2}}{f_1\sqrt{u_x}u_y}w_{22,y}
    + \frac{2f_{3, t}u_tu_y + 2f_3u_tu_{yt} + f_3u_yu_{tt}}{2f_3u_tu_y}w_{23}
    + \frac{4f_3\sqrt{u_t}}{u_y}p_y.
\end{align*}

\subsection{$A = -2C$, $B = -2C$}\label{sec:a-=-2c-1}

We have
\begin{align*}
  \mathcal{R}_2
  &= \left(2f_3u_t^2u_{tt} - \frac{f_3u_t^3u_{xt}}{u_x} + f_{3,
    t}u_t^3\right)q
    + \frac{f_3u_t^4}{2u_{xx}}q_x\\
  &-  f_1u_x^{3/2}\left(f_3u_t^2u_xu_{tt} - \frac{f_3u_t^3u_{xt}}{4}
    + \frac{f_{3, t}u_xu_t^3}{2}\right)w_1 + u_tw_{31,t},\\
  \mathcal{R}_3
  &= \nu_3q + \left(\frac{5u_t^3u_tt}{4u_x} - \frac{3u_t^4u_{xt}}{8u_x^2} +
    \frac{f_{3, t}u_t^4}{2f_3u_x}\right)q_x
    - \frac{u_t^3u_{xt}}{4u_x}q_t\\
  &+
    \frac{u_t^4}{4u_x}q_{xt} - \mu_3w_1 + \frac{u_t}{2f_3}w_{31,tt},
    \intertext{where}
    \nu_3
  &= -\frac{u_t^3u_{xtt}}{2u_x} + u_t^2u_{ttt}
    + \frac{3u_t^3u_{xt}^2}{4u_x^2} - \frac{(f_{3, t}u_t^2u_t +
    2f_3u_{tt})u_{xt}}{f_3u_x} + u_t u_{tt}^2\\
  &+ \frac{2f_{3, t}u_t^2u_{tt}}{f_3} + \frac{f_{3, t, t}u_t^3}{2f_3},\\
  \mu_3
  &= -\frac{u_t^3u_{xtt}}{8f_1u_x^{3/2}}
    + \frac{u_t^2u_{ttt}}{2f_1\sqrt{u_x}}
    + \frac{3u_t^3u_{xt}^2}{16f_1u_x^{5/2}}
    - \frac{u_t^2(f_{3, t}u_t + 2f_3u_{tt})u_{xt}}{4f_1f_3u_x^{3/2}}\\
  &+ \frac{u_tu_{tt}^2}{2f_1\sqrt{u_x}}
    +\frac{f_{3, t}u_t^2u_{tt}}{f_1f_3\sqrt{u_x}}
    + \frac{f_{3, t, t}u_t^3}{4f_1f_3\sqrt{u_x}}.
\end{align*}
We also have
\begin{align*}
  w_{1,y}
  &= \frac{4f_1(u_xu_{yt} + u_yu_{xt})}{\sqrt{u_x}u_t}q
    - \frac{2f_1u_y}{\sqrt{u_x}}q_x,\\
  w_{1,t}
  &= \frac{2f_1u_{xt}}{\sqrt{u_x}}q
    - \frac{f_1u_t}{\sqrt{u_x}}q_x + \frac{2f_1}{\sqrt{u_x}}q_t,\\
  w_{31,x}
  &= -2f_3u_tu_{xt}q - f_3u_t^2q_x
    - f_3\left(\frac{u_t^2u_{xx}}{4f_1u_x^{3/2}}
    - \frac{u_tu_{xt}}{f_1\sqrt{u_x}}
    + \frac{f_{1, x}u_t^2}{2f_1^2\sqrt{u_x}}\right)w_1\\
  &+ \frac{f_3u_t^2}{2f_1\sqrt{u_x}}w_{1,x},\\
  w_{31,y}
  &= \frac{2f_3u_tu_yu_{xt}}{u_x}q - \frac{f_3u_t^2u_y}{u_x}q_x - f_3u_t^2q_y
    + \frac{f_3u_t(u_xu_{yt} - u_yu_{xt})}{2f_1u_x^{3/2}}w_1
  \intertext{and}
    w_{32,x}
  &= -\frac{2f_2\sqrt{u_y}(u_tu_{xt} - 2u_xu_{tt})}{u_t^3}p
    - \frac{4f_2\sqrt{u_y}u_x}{u_t^2}p_t
    + \frac{u_xu_{yt} - u_yu_{xt}}{u_tu_y}w_{32},\\
  w_{32,y}
  &= -\frac{4f_2u_y^{3/2}u_{tt}}{u_t^3}p - \frac{2f_2\sqrt{u_y}}{u_t}p_y
    + \frac{4f_2u_y^{3/2}}{u_t^2}p_t\\
  &+ \left(\frac{u_{yy}}{2u_y} - \frac{2u_{yt}}{u_t}
    + \frac{f_{2, y}}{f_2}\right)w_{32},\\
  w_{33,x}
  &= \frac{8f_1\sqrt{u_x}(u_tu_{xt} - u_xu_{tt})}{u_t^3}p
    - \frac{2f_1\sqrt{u_x}}{u_t}p_x
    + \frac{8f_1u_x^{3/2}}{u_t^2}p_t\\
  &- \frac{f_1\sqrt{u_x}(u_xu_{yt} - 4u_yu_{xt})}{f_2u_tu_y^{3/2}}w_{32}
    + \frac{2f_1u_x^{3/2}}{u_tf_2\sqrt{u_y}}w_{32,t}\\
  &+ \left(\frac{u_{xx}}{2u_x} + \frac{u_{xt}}{u_t}
    + \frac{f_{1, x}}{f_1}\right)w_{33},\\
  w_{33,y}
  &= \frac{4f_1\sqrt{u_x}(u_tu_{yt} + u_yu_{tt})}{u_t^3}p
    - \frac{4f_1u_y\sqrt{u_x}}{u_t^2}p_t
    + \frac{3f_1\sqrt{u_x}u_{yt}}{f_2u_t\sqrt{u_y}}w_{32}\\
  &- \frac{2f_1\sqrt{u_x}\sqrt{u_y}}{f_2u_t}w_{32,t}
    + \frac{2u_xu_{yt} + u_yu_{xt}}{u_tu_x}w_{33}.
\end{align*}

\section{Nonlocal variables of the \ABC-equation}
\label{sec:nonl-vari-abc}

The nonlocal variables associated with the two-component conservation laws
read as follows:
\subsection{$A + B + C = 0$}
\label{sec:+-b-+}
\begin{align*}
  z_{1,y}&= \frac{(a - 1)f_1u_y}{u_x},\quad
  z_{1,t}= -\frac{f_1u_t}{u_x};\\
  z_{2,x}&= \frac{af_2u_x}{u_y},\quad
  z_{2,t}= \frac{f_2u_t}{u_y};\\
  z_{3,x}&= \frac{af_3u_x}{u_t},\quad
  z_{3,y}= \frac{(a - 1)f_3u_y}{u_t}.
\end{align*}

\subsection{$A = -C, B = -C$}
\label{sec:=-c-b}
\begin{align*}
  z_{11,y}
  &= \frac{f_1uu_yu_{xx}}{2u_x}
    + \frac{f_1u\ln{(u_x)}u_yu_{xt}}{2u_t}
    + \frac{f_1 u\ln{(u_x)}u_xu_{yt}}{2u_t}
    + \frac{f_1 (\ln{(u_x)} - 3) u_y u_x}{2},\\
  z_{11,t}
  &= \frac{f_1 uu_tu_{xx}}{2u_x}
    + \frac{f_1 u (\ln{(u_x)} - 1) u_{xt}}{2}
    + \frac{f_1u_t(\ln{(u_x)} - 2) u_x}{2};\\
  z_{12,y}
  &= -f_1 u_y \ln{(u_x)},\quad
  z_{12,t}
  = -f_1 (\ln{(u_x)} - 1) u_t;\\
  z_{21,x}
  &= -f_2 u_x \ln{(u_y)},\quad
  z_{21,t}
  = -f_2 (\ln{(u_y)} - 1) u_t;\\
  z_{22,x}
  &= \frac{f_2u\ln{(u_y)}u_y u_{xt}}{u_t}
    + \frac{f_2 u u_x u_{yy}}{u_y}
    + \frac{f_2 uu_x\ln{(u_y)} u_{yt}}{u_t}
    + f_2 u_x u_y (\ln{(u_y)} - 3),\\
  z_{22,t}
  &= \frac{f_2 u u_t u_{yy}}{u_y}
    + f_2u (\ln{(u_y)} - 1)u_{yt} + f_2u_t u_y (\ln{(u_y)} - 2);\\
  z_{3,x}
  &= \frac{f_3 u_x}{2u_t^2},\quad
  z_{3,y}
  = -\frac{f_3 u_y}{2u_t^2}.
  \end{align*}

\subsection{$A = -C, B = C$}
\label{sec&=-c-b-1}
\begin{align*}
  z_{1,y}
  &= -\frac{f_1 u_y}{2u_x^2},\quad
  z_{1,t}
  = \frac{f_1u_t}{2u_x^2};\\
  z_{21,x}
  &= -f_2 u_x \ln{(u_y)},\quad
  z_{21,t}
  = -f_2 (\ln(u_y) + 1) u_t;\\
  z_{22,x}
  &= -\frac{f_2u\ln{(u_y)}u_y u_{xt}}{u_t}
    - f_2 (1 + \ln{(u_y)})u_y u_x + \frac{f_2 u u_x u_{yy}}{u_y}
    + \frac{f_2 u_x \ln{(u_y)} u u_{yt}}{u_t},\\
  z_{22,t}
  &= \frac{f_2 u u_t u_{yy}}{u_y}
    + f_2 u (\ln(u_y) + 1) u_{yt} + f_2u_t u_y (\ln{(u_y)} - 2);\\
  z_{31,x}
  &= -f_3 u_x (\ln{(u_t)} - 1),\quad
  z_{31,y}
  = -f_3 \ln{(u_t)} u_y;\\
  z_{32,x}
  &= \frac{f_3 u u_x u_{tt}}{2 u_t}
    + \frac{f_3 u \ln{(u_t)}u_xt}{2}
    + \frac{u_x (f_3 u_t \ln{(u_t)} + f_3'u - f_3 u_t)}{2},\\
  z_{32,y}
  &= \frac{f_3 u u_y u_{tt}}{2u_t}
    + \frac{f_3u(\ln{(u_t)} + 1) u_yt}{2} +
    \frac{u_y (f_3 u_t \ln{(u_t)} + f_3'u - 2 f_3 u_t)}{2}.
\end{align*}

\subsection{$A = -C, 2B = -3C$}
\label{sec&=-c-2b}
\begin{align*}
  z_{1,y}
  &= -3 f_1 \sqrt{u_x} u_y,\quad
  z_{1,t}
  = -2 f_1 \sqrt{u_x} u_t;\\
  z_{21,x}
  &= \frac{3f_2u\ln{(u_y)}u_yu_{xt}}{4u_t}
    - \frac{7f_2 u_x u_y}{4}
    + \frac{f_2u_y\ln{(u_y)} u_x}{2}\\
  &+ \frac{f_2uu_x \ln{(u_y)}u_{yt}}{2u_t}
    + \frac{f_2 u u_x u_{yy}}{2u_y},\\
  z_{21,t}
  &= \frac{f_2 u u_t u_{yy}}{2 u_y}
    + \frac{f_2u (2 \ln{(u_y)} - 3)u_{yt}}{4}
    + \frac{f_2u_t u_y (\ln{(u_y)} - 2)}{2};\\
  z_{22,x}
  &= -\frac{f_2 (2 \ln{(u_y)} + 3) u_x}{2},\quad
  z_{22,t}
  = -f_2 u_t \ln{(u_y)};\\
  z_{3,x}
  &= \frac{2 f_3 u_x}{u_t^{5/2}},\quad
  z_{3,y} = -\frac{3 f_3 u_y}{u_t^{5/2}}.
\end{align*}

\subsection{$A = -C, B = -3C$}
\label{sec&=-c-b-2}
\begin{align*}
  z_{1,y}
  &= -3 f_1 u_x^2 u_y,\quad
  z_{1,t}
  = -f_1 u_x^2 u_t;\\
  z_{21,x}
  &= \frac{3f_2u\ln{(u_y)}u_y u_{xt}}{2u_t}
    - \frac{5 f_2 u_x u_y}{2}
    + \frac{f_2 u u_x u_{yy}}{2u_y}\\
  &+ \frac{f_2 u_y \ln{(u_y)} u_x}{2}
    + \frac{f_2u u_x \ln{(u_y)}u_{yt}}{2u_t},\\
  z_{21,t}
  &= \frac{f_2 u u_t u_{yy}}{2u_y}
    +\frac{f_2 u (\ln{(u_y)} - 3)u_{yt}}{2}
    +\frac{f_2u_t u_y (\ln{(u_y)} - 2)}{2};\\
  z_{22,x}
  &= -f_2 u_x \ln{(u_y)},\quad
  z_{22,t}
  = -f_2 u_t (\ln{(u_y)} - 3);\\
  z_{3,x}
  &= \frac{f_3 u_x}{u_t^4},\quad
  z_{3,y}
  = -\frac{3 f_3 u_y}{u_t^4}.
\end{align*}

\subsection{$A = -C, 2B = 3C$}
\label{sec&=-c-2b-1}
\begin{align*}
  z_{1,y}
  &= -\frac{3 f_1 u_y}{u_x^{5/2}},\quad
  z_{1,t}
  = \frac{2 f_1 u_t}{u_x^{5/2}};\\
  z_{21,x}
  &= -\frac{3 f_2 u\ln{(u_y)}u_y u_{xt}}{4u_t}
    - \frac{f_2 u_x u_y}{4}
    + \frac{f_2 u u_x u_{yy}}{2u_y}\\
  &+ \frac{f_2 u_y \ln{(u_y)} u_x}{2}
    + \frac{f_2 uu_x \ln{(u_y)}u_{yt}}{2 u_t},\\
  z_{21,t}
  &= \frac{f_2 u u_t u_{yy}}{2 u_y}
    + \frac{f_2 u (2 \ln{(u_y)} + 3) u_yt}{4}
    + \frac{f_2 u_t u_y (\ln{(u_y)} - 2)}{2};\\
  z_{22,x}
  &= -f_2 u_x \ln{(u_y)},\quad
  z_{22,t}
  = -\frac{f_2 u_t (2 \ln{(u_y)} + 3)}{2};\\
  z_{3,x}
  &= -2f_3 \sqrt{u_t}u_x,\quad
  z_{3,y}
  = -3f_3 \sqrt{u_t}u_y.
\end{align*}

\subsection{$A = -C, B = 3C$}
\label{sec&=-c-b-3}
\begin{align*}
  z_{1,y}
  &= -\frac{3 f_1 u_y}{u_x^4},\quad
  z_{1,t}
  = \frac{f_1 u_t}{u_x^4};\\
  z_{2,x}
  &= -f_2 u_x \ln{(u_y)},\quad
  z_{2,t}
  = -f_2 u_t (\ln{(u_y)} + 3);\\
  z_{3,x}
  &= -f_3 u_x u_t^2,\quad
  z_{3,y}
  = -3 f_3 u_y u_t^2.
\end{align*}

\subsection{$3A = 2C, 3B = -2C$}
\label{sec&3a-=-2c-2}
\begin{align*}
  z_{1,y}
  &= -\frac{2 f_1 \sqrt{u_x} u_y}{3},\quad
  z_{1,t}
  = -f_1 \sqrt{u_x} u_t;\\
  z_{2,x}
  &= -\frac{2 f_2 u_x}{3 u_y^{5/2}},\quad
  z_{2,t}
  = \frac{f_2 u_t}{u_y^{5/2}};\\
  z_{31,x}
  &= f_3 u_x (2 \ln{(u_t)} + 3),\quad
  z_{31,y}
  = 2 f_3 \ln{(u_t)} u_y;\\
  z_{32,x}
  &= \frac{7 f_3 u_x u_t}{6}
    + \frac{f_3'u u_x}{2}
    - \frac{f_3 u_t u_x \ln{(u_t)}}{3}
    - \frac{f_3 u\ln{(u_t)}u_{xt}}{3}
    - \frac{f_3 u u_x u_{tt}}{3 u_t},\\
  z_{32,y}
  &= -\frac{f_3 u u_y u_{tt}}{3 u_t}
    - \frac{f_3u(2 \ln{(u_t)} - 3) u_yt}{6}
    + \frac{f_3'u u_y}{2} - \frac{f_3u_y(\ln{(u_t)} - 2) u_t}{3}.
\end{align*}

\subsection{$3A = C, 3B = -C$}
\label{sec&3a-=-c-2}
\begin{align*}
  z_{1,y}
  &= -\frac{f_1 u_y}{12 u_x^4},\quad
  z_{1,t}
  = \frac{f_1 u_t}{4 u_x^4};\\
  z_{2,x}
  &= -\frac{f_2 u_x u_y^2}{6},\quad
  z_{2,t}
  = -\frac{f_2 u_t u_y^2}{2};\\
  z_{31,x}
  &= -\frac{f_3 (\ln{(u_t)} - 3) u_x}{3},\quad
  z_{31,y}
  = -\frac{f_3 \ln{(u_t)} u_y}{3};\\
  z_{32,x}
  &= \frac{f_3 u u_x u_{tt}}{6 u_t}
    + \frac{f_3 u u_{xt} \ln(u_t)}{6}
    + \frac{u_x (f_3 u_t \ln(u_t) + 3f_3'u + f_3 u_t)}{6},\\
  z_{32,y}
  &= \frac{f_3 u u_y u_{tt}}{6 u_t}
    + \frac{f_3 u (3 + \ln{(u_t)}) u_{yt}}{6}
    + \frac{u_y (f_3 u_t \ln{(u_t)} + 3f_3'u - 2 f_3 u_t)}{6}.
  \end{align*}

\subsection{$A = C, 2B = -C$}
\label{sec&=-c-2b-2}
\begin{align*}
  z_{1,y} &= -f_1 \sqrt{u_x} u_y,\quad
  z_{1,t} = -2 f_1 \sqrt{u_x} u_t;\\
  z_{2,x} &= -\frac{f_2 u_x}{u_y^4},\quad
  z_{2,t} = \frac{f_2 u_t}{u_y^4};\\
  z_{3,x} &= 2f_3 \sqrt{u_t}u_x,\quad
  z_{3,y} = f_3\sqrt{u_t}u_y.
\end{align*}

\subsection{$A = -2C, B = -2C$}
\label{sec&=-2c-b}
\begin{align*}
  z_{1,y} &= -2 f_1 \sqrt{u_x} u_y,\quad
  z_{1,t} = -f_1 \sqrt{u_x} u_t;\\
  z_{2,x} &= -2f_2 \sqrt{u_y}u_x,\quad
  z_{2,t} = -f_2\sqrt{u_y}u_t;\\
  z_{3,x} &= \frac{f_3 u_x}{u_t^4},\quad
  z_{3,y} = -\frac{2f_3u_y}{2u_t^4}.
\end{align*}

\section{Nonlocal cosymmetries}
\label{sec:actions-1}

\subsection{The case $A = -C$, $B = -C$}
\label{sec:case-a-=}
We have:
\begin{align*}
  \Psi_2
  &= -\frac{4u_tz_{11}}{f_1u_x}
    - \frac{(2f_1u_x - f'u)u_tz_{12}}{f_1^2u_x}
    - \frac{uu_tz_{12,x}}{f_1u_x}\\
  &- \frac{(2f_2u_y - f'u) u_t z_{21}}{f_2^2u_y} -
    \frac{uu_tz_{21,y}}{f_2u_y}
    - \frac{2u_tz_{22}}{f_2u_y} - u (\ln(u_x) + \ln(u_y) - 4) u_t,\\
  \Psi_3
  &= \frac{(2f'g_1 - f_1g')u_t z_{11}}{f_1^2u_x} -
    \frac{2g_1u_tz_{11,x}}{f_1u_x}\\
  &+ \frac{(f_1g'f'u
    - 2f'^2g_1u + 2f_1g_1f'u_x
    + f_1''f_1 g_1 u) u_t z_{12}}{f_1^3 u_x}\\
  &- \frac{(f_1g_1'u - 2g_1f_1'u + 2f_1g_1u_x)u_tz_{12,x}}{f_1^2u_x}
    - \frac{g_1uu_tz_{12,xx}}{f_1u_x}\\
  &-
    \frac{(2g_1\ln(u_x)u_x^2 - 6g_1 u_x^2 - 4g_1'uu_x
    - g_1''u^2) u_t}{2u_x},\\
  \Psi_4
  &= \frac{(f_2''f_2g_2 uu_t - 2f_2'^2 g_2 u
    + f_2'g_2'f_2 u + 2g_2f_2'f_2 u_y)z_{21}}{f_2^3u_y}\\
  &- \frac{u_t(2g_2f_2u_y - 2f_2'g_2 u + g_2'f_2 u)z_{21,y}}{f_2^2u_y}
    - \frac{g_2uu_tz_{21,yy}}{f_2u_y}\\
  &+ \frac{(g_2f_2' - f_2g_2')u_tz_{22}}{f_2^2u_y}
    - \frac{g_2 u_t z_{22,y}}{f_2u_y}
    + \frac{g_2''u^2u_t}{2u_y} + 2g_2'u_t u - g_2u_y(\ln u_y - 3) u_t
\end{align*}

\subsection{Case $A = -C$, $B = C$}
\label{sec:nonl-cosymm-case}

We have
\begin{align*}
  \Psi_1
  &= \frac{f_2'u_tz_{21}}{f_2^2u_y}
    - \frac{u_tz_{21,y}}{f_2u_y} +
    \frac{f_3'z_{31}}{f_3^2} + \frac{z_{31,t}}{f_3}
    + u_t (\ln(u_y) + \ln(u_t) - 1),\\
  \Psi_2
  &= \frac{u_t(2f_2u_y - f_2'u)z_{21}}{f_2^2u_y}
    + \frac{uu_tz_{21,y}}{f_2u_y}
    + \frac{2u_tz_{22}}{f_2u_y}
    + \frac{(2f_3u_t - f_3'u)z_{31}}{f_3^2}\\
  &+ \frac{uz_{31,t}}{f_3}
    + \frac{4z_{32}}{f_3}
    + \frac{u(f_3u_t\ln(u_t) + f_3u_t\ln(u_y) - 5f_3u_t - f_3'u)}{f_3},
  \\
  \Psi_3
  &=\frac{u_t(f_2g_2f_2'' u - 2f_2'^2g_2u  +
    2f_2 f_2'g_2u_y + f_2f_2'g_2'u)z_{21}}{f_2^3u_y}\\
  &- \frac{(2f_2g_2u_y - 2g_2f_2'u + f_2g_2'u)u_tz_{21,y}}{f_2^2u_y}
    - \frac{g_2uu_tz_{21,yy}}{f_2u_y}\\
  &+\frac{(f_2'g_2 - f_2g_2')u_tz_{22}}{f_2^2u_y}
    - \frac{g_2u_tz_{22,y}}{f_2u_y}\\
  &- \frac{u_t (2g_2u_y^2\ln(u_y) -
  4g_2u_y^2 - 4g_2'uu_y - g_2''u^2)}{2u_y}\\
  \Psi_4
  &= \frac{(2f_3'f_3g_3u_t - 2f_3'^2g_3u
    + f_3'g_3'f_3u + f_3''f_3 g_3 u) z_{31}}{f_3^3}\\
  &- \frac{(2f_3g_3u_t - 2f_3'g_3u + f_3g_3u)z_{31,t}}{f_3^2}
    - \frac{g_3uz_{31,tt}}{f_3}
    + \frac{2(g_3f_3' - f_3g_3')z_{32}}{f_3^2}\\
  &- \frac{2g_3z_{32,t}}{f_3}
    - \frac{(f_3^2g_3 - f_3g_3f_3
    - f_3''f_3g_3 - g_3''f_3^2)u^2}{2f_3^2}\\
  &+ \frac{(f_3'g_3u_t + 3f_3g_3'u_t + f_3g_3u_{tt})u}{f_3}
    - g_3u_t^2(\ln(u_t) - 4)
\end{align*}

\subsection{$A = -C$, $2B = -3C$}
\label{sec:nonl-cosymm-a}
One has
\begin{align*}
  \Psi_{1,2}^1
  &= \frac{3u_tz_1}{2f_1\sqrt{u_x}}
    + \frac{16 u_t z_{21}}{f_2u_y} +
    \frac{4u_t(2f_2u_y - f_2'u)z_{22}}{f_2^2u_y}\\
  &+
  \frac{4 u u_t z_{22,y}}{f_2u_y} + (4 u (\ln(u_y) + 6) u_t,\\
  \Psi_2^2
  &= -\frac{8 u_t(g_2 f_2' - f_2 g_2')z_{21}}{f_2^2u_y}
  + \frac{8g_2u_tz_{21,y}}{f_2u_y}\\
  &+ \frac{4u_t(2f_2'^2g_2u - g_2'f_2'f_2u
    - 2f_2'f_2g_2 u_y - f_2''f_2 g_2 u) z_{22}}{f_2^3u_y }\\
  & - \frac{4u_t(2f_2'g_2u - g_2'f_2u - 2f_2g_2 u_y)z_{22,y}}{f_2^2u_y}
    + \frac{4g_2 uu_t z_{22,yy}}{f_2u_y}
    + \frac{2g_2 uu_t u_{yy}}{3u_y}\\
  &- \frac{(-4g_2u_y^2\ln(u_y) + 2g_2'u u_y + 2g_2''u^2 + g_2u_y^2)u_t}{u_y}.
\end{align*}

\end{document}